\documentclass[aps,prl,reprint,groupedaddress,showpacs]{revtex4-1}
\usepackage{graphicx}
\usepackage{latexsym,amssymb,amsmath}
\usepackage{bm}
\usepackage[caption=false]{subfig}
\usepackage{xcolor}

\usepackage[colorlinks=true,
urlcolor=blue,
linkcolor=blue,
citecolor=blue
]{hyperref}
\begin{document}

%\preprint{}

\title{Role of Near-Field Interaction on Light Transport in Disordered Media}
\author{B. X. Wang}
\affiliation{Institute of Engineering Thermophysics, Shanghai Jiao Tong University, Shanghai, 200240, P. R. China}


\author{C. Y. Zhao}
\email{Changying.zhao@sjtu.edu.cn}
%\thanks{}
\affiliation{Institute of Engineering Thermophysics, Shanghai Jiao Tong University, Shanghai, 200240, P. R. China}


\date{\today}

\begin{abstract}
Understanding light-matter interaction in disordered photonic media allows people to manipulate light scattering and achieve exciting applications using seemingly scrambled media. As the concentration of scattering particles rises, they are inclined to step into near fields of each other in deep subwavelength scale. The fundamental physics involving the interplay between disorder and near-field interaction (NFI) is still not fully understood. We theoretically examine the role of NFI by analyzing the underlying multiple scattering mechanism. We find NFI leads to a stronger collective behavior involving more particles and widens the photonic pseudo-bandgap of disordered media. It also excites more weakly decayed longitudinal modes and results in higher local density of states. By introducing a sticky short-range order, we demonstrate the possibility of enhancing off-momentum-shell NFI of multiple scattering process. Our results have profound implications in understanding and harnessing nanoscale light-matter interaction for novel disordered photonic devices. 
\end{abstract}


%\keywords{}

\maketitle

%\section{}
%\subsection{}
%\subsubsection{}

\textit{Introduction}.-- When light propagates in disordered photonic media, it undergoes scattering in a very complicated way. The large coherence time, elastic scattering and time-reversible characteristics of light guarantee the interference phenomena in mescoscopic and even macroscopic scales. Rich interference phenomena in disordered media have received growing attention in the last a few years, and give rise to a rapidly developing field called ``disordered photonics'' \cite{wiersma2013disordered,Rotter2017}. Most of these studies in this field are theoretically investigated by means of random matrix theory (RMT) \cite{Beenakker1997} or multiple scattering theory (MST) for scalar waves \cite{sheng2006introduction}. As the concentration of scattering particles in disordered media rises, they are inclined to step into the near fields of each other \cite{Liew2011,Naraghi2015}, making the far-field and on-shell approximations invalid \cite{wangPRA2018}, because near-field interaction (NFI) will introduce electromagnetic wave components with large wavevectors \cite{wangPRA2018}. These components, that decay very fast in the far-field region, may contribute to radiation energy tunneling and thus transport properties \cite{Naraghi2015}. Moreover, Near-field interaction (NFI) among scatterers is purely vectorial containing both transverse and longitudinal components \cite{lagendijk1996resonant,aubryPRA2017}, leading to a more intriguing picture of many-body coherent scattering, which MST for scalar waves is not able to fully capture. RMT also lacks the capability to directly distinguish NFI. The role of NFI, as a fundamental mechanism, has not been thoroughly studied yet. NFI between closely packed random scatterers brings rich phenomena like inducing photon tunneling \cite{Naraghi2015} and ``phase transition" \cite{Naraghi2016}, leading to unusual structural colors \cite{Liew2011} and affecting quantum emission \cite{Sapienza2011}. It was also confirmed to enhance total transmission of disordered media \cite{Naraghi2015}, as well as demonstrated being a hindering factor for Anderson localization in three-dimension \cite{Skipetrov2014,escalanteADP2017}, while Silies et al. recently found that near-field coupling surprisingly assists the formation of localized modes \cite{Silies2016}. Pierrat and co-workers reported that near-field interaction of a dipolar emitter with more than one particle creates optical modes confined in a small volume around it and give rise to strong fluctuations in LDOS \cite{Pierrat2010}.  Nevertheless these studies still lack an explicit demonstration for the role of near-field interaction and the underlying physical mechanism. Here we theoretically study the role of NFI on light scattering and transport in disordered media, by addressing how NFI affects the many-body scattering mechanism and its interplay with short-range order, which then allows a flexible control over light-matter interaction in random media. Since NFI is much stronger than far-field interaction, it is promising to utilize NFI to achieve extreme light-matter interaction, facilitating the performance of novel photonic devices through modifying NFI using short-range order. 
%
%The study of disordered photonics includes the fundamental pursuit of Anderson localization of light of photons in various disordered micro/nanostructures \cite{wiersma1997localization,Segev2013,Sperling2016NJP} and cold atomic clouds \cite{Skipetrov2014,Skipetrov2015}, focusing \cite{Vellekoop2010} and imaging \cite{Mosk2012}, random lasers \cite{Cao1999,Wiersma2008}, amorphous photonic crystals \cite{Florescu2009,Froufe-Perez2016}, and photovoltaics \cite{Vynck2012}, etc. Statistics \cite{Chabanov2000}, spatial and spectral correlations \cite{Dogariu2015} as well as coherent control for light waves \cite{Rotter2017,hsu2017correlation} were also investigated extensively.
%The traditional picture of light intensity transport is depicted by radiative transfer equation (RTE) \cite{lagendijk1996resonant,VanRossum1998,tsang2004scattering,sheng2006introduction,akkermans2007mesoscopic}. Parameters describing radiative transport are usually calculated under independent scattering approximation (ISA), i.e., in which the scatterers scatter electromagnetic (EM) waves independently without any inter-particle interference taken into account \cite{lagendijk1996resonant,VanRossum1998,tsang2004scattering,sheng2006introduction,akkermans2007mesoscopic}. However,
%In quantum optics, NFI among an ensemble of atoms induces exciting collective and cooperative phenomena including Dicke’s superradiance and subradiance \cite{Rohlsberger2010}.

\textit{Model.}--A particle’s scattering field, e.g., a point dipole, shows very different behaviors in its near field and far field, and is expressed as $\mathbf{E}_s(\omega,\mathbf{r})=\mathbf{G}_0(\omega,\mathbf{r})\mathbf{d}(\omega)$, where $\mathbf{d}(\omega)$  is the dipole moment, and $\mathbf{G}_0(\omega,\mathbf{r})$ is free-space dyadic Green's tensor containing both near-field and far-field scattered wave components.
%\begin{equation}
%\begin{split}
%\mathbf{G}_{0}(\omega,\mathbf{r})=&-(\frac{i}{kr}-\frac{1}{(kr)^2}+1)\frac{\exp{(ikr)}}{4\pi %r}(\mathbf{I}-\mathbf{\hat{r}}\mathbf{\hat{r}})\\
%&+2(\frac{i}{kr}-\frac{1}{(kr)^2})\frac{\exp{(ikr)}}{4\pi %r}\mathbf{\hat{r}}\mathbf{\hat{r}}+\frac{\delta(\mathbf{r})}{3k^{2}}\mathbf{I}
%\end{split}
%\end{equation}
The near field contains fast-decaying components varying with radial distance $r$ from the dipole as $1/r^3$ and $1/r^2$ respectively, which originate from the electrostatic Coulomb field of the dipole and has both longitudinal and transverse components, while the nature of scattering far field is retardation effect obeying a $1/r$ law with only a transverse field \cite{lagendijk1996resonant}. When many such particles are brought together randomly to build up a disordered medium, the near-field coupling between individual scattering fields strongly affects the whole medium's EM response. Without loss of physical significance, we consider a disordered medium composed of point dipole scatterers with a polarizability of $\alpha(\omega)=-{3\pi\ c^3\gamma}/{[\omega^3(\omega-\omega_0 + i \gamma/2)]}$ with a resonance at ${\omega_{0}=3.42\times 10^{15} \mathrm{rad/s}}$   and a line-width of ${\gamma=1\times 10^{9} \mathrm{rad/s}}$, aiming to investigate the collective resonant behavior \cite{aubryPRA2017}. To describe particle correlations in typical disordered materials, an artificial spherical radius of $a$ is assigned to each scatterer, which also defines particle volume fraction as $f_v=4\pi n_0a^3/3$ with a volume-averaged number density of $n_0$. In the general case of an infinite 3D medium we can write down Dyson`s equation in Fourier space for configurational averaged, retarded amplitude Green`s tensor when assuming the disordered medium is isotropic and translational-invariant as $\mathbf{G}_{c}^{-1}(\omega,\mathbf{p})={\mathbf{G}_{0}^{-1}(\omega,\mathbf{p})-\mathbf{\Sigma}(\omega,\mathbf{p})}$, where $ \mathbf{G}_{0}(\omega,\mathbf{p})=\left[k^{2}\mathbf{I}-p^{2}(\mathbf{I}-\mathbf{\hat{p}}\mathbf{\hat{p}})\right]^{-1}$ is the bare Green'{s} tensor in vacuum fulfilling the vector Helmholtz equation for electromagnetic waves \cite{lagendijk1996resonant,VanRossum1998}. Here $\mathbf{\hat{p}}$ is the unit vector in the momentum space, and $k=\omega/c$ is the wave number in vacuum. $\mathbf{\Sigma}(\omega,\mathbf{p})$ is the self energy tensor which provides a renormalization of the disordered media, and determines the effective (renormalized) permittivity as $\bm{\varepsilon}(\omega,\mathbf{p})=\mathbf{I}-\mathbf{\Sigma}(\omega,\mathbf{p})/k^{2}$ \cite{lagendijk1996resonant}. The obtained momentum-dependent effective permittivity tensor is decomposed into a transverse part and a longitudinal part as $
\bm{\varepsilon}(\omega,\mathbf{p})=\varepsilon^{\bot}(\omega,\mathbf{p})(\mathbf{I}-\mathbf{\hat{p}}\mathbf{\hat{p}})+\varepsilon^{\parallel}(\omega,\mathbf{p})\mathbf{\hat{p}}\mathbf{\hat{p}}$, where $\varepsilon^{\bot}(\omega,\mathbf{p})=1-\Sigma^{\bot}(\omega,\mathbf{p})/{k^{2}}$ and $\varepsilon^{\parallel}(\omega,\mathbf{p})=1-\Sigma^{\parallel}(\omega,\mathbf{p})/{k^{2}}$ determine the effective permittivities of transverse and longitudinal modes in momentum space. Therefore, by determining the poles of amplitude Green{'}s function we can obtain the dispersion relation which corresponds to collective excitation of the disordered medium. In practice, this amounts to find the maxima of spectral function defined as $\mathbf{S(\omega,\mathbf{p})}=-\mathrm{Im}{\mathbf{G}_c(\omega,\mathbf{p})}$ \cite{sheng2006introduction}.

\begin{figure}
	\centering
	\includegraphics[width=0.8\linewidth]{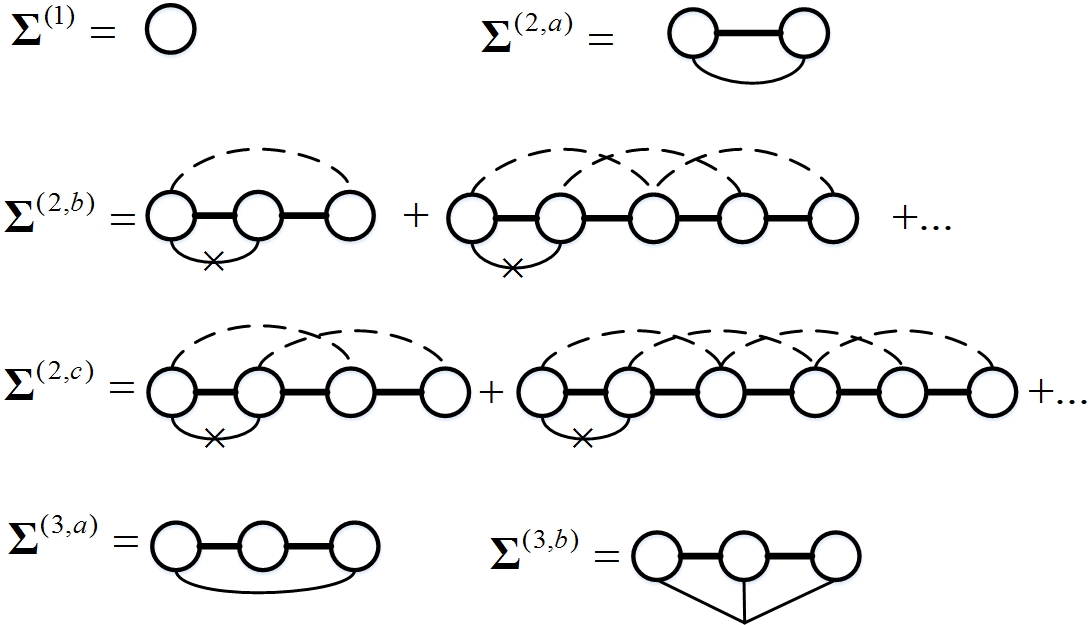}
	\caption{The perturbative diagrammatic expansion for self-energy considering various many-particle mechanisms. Circles denote particles described by a t-matrix $\mathbf{T}=-k^2\alpha\mathbf{I}$; Thin solid lines denote two-particle or three-particle correlation functions $h_2(\mathbf{r}_1,\mathbf{r}_2)$ or $h_3(\mathbf{r}_1,\mathbf{r}_2,\mathbf{r}_3)$; Thin solid lines with a cross denote the pair distribution function and $g_2(\mathbf{r}_1,\mathbf{r}_2)=1+h_2(\mathbf{r}_1,\mathbf{r}_2)$; Bold solid lines denote free-space Green's tensor $\mathbf{G}_0(\omega,\mathbf{r})$; Dashed lines denote the same particle \cite{sm}.}\label{fig1c}
\end{figure}

%By finding the maxima of spectral function in real momentum space, the transverse (propagating) modes with a real part of wavenumber satisfying $K^{\bot 2}=\mathrm{Re}\varepsilon^{\bot}(\omega,K^{\bot})k_{0}^{2} $ are identified while keeping the imaginary parts of effective permittivity as small as possible, with a meaning of “most probable propagating wave” \cite{sheng2006introduction,Page1996}. 

%For transverse excitation, the dispersion relation is $\mathbf{K}^{\bot 2}=\mathbf{\varepsilon}^{\bot}(\omega,\mathbf{K^{\bot}})k^{2}$. $\mathbf{K}^{\bot} $ is viewed as the effective propagation wave vector for the disordered medium. On the other hand, if for some  $\mathbf{K}^{\parallel}$, $\mathbf{\varepsilon}^{\parallel}(\omega,\mathbf{K^{\parallel}})k^{2}=0$ is fulfilled, non-propagating longitudinal modes are also supported. 

%\begin{equation}
%\begin{split}
%\mathbf{S(\omega,\mathbf{p})}=&-\mathrm{Im}{\mathbf{G}_c(\omega,\mathbf{p})}=\frac{\mathrm{Im}\varepsilon^{\parallel}(\omega,\mathbf{p})/k_0^2}{[\mathrm{Re}\varepsilon^{\parallel}(\omega,\mathbf{p})]^2+[\mathrm{Im}\varepsilon^{\parallel}(\omega,\mathbf{p})]^2}\mathbf{\hat{p}}\mathbf{\hat{p}}\\
%&+\frac{\mathrm{Im}\varepsilon^{\bot}(\omega,\mathbf{p})/k^2}{[\mathrm{Re}\varepsilon^{\bot}(\omega,\mathbf{p})-p^%2/k^2]^2+[\mathrm{Im}\varepsilon^{\bot}(\omega,\mathbf{p})]^2}(\mathbf{I}-\mathbf{\hat{p}}\mathbf{\hat{p}})
%\end{split}
%\end{equation}

We calculated self-energy in a perturbative way and derived analytical formulations after recognizing critical multiple scattering mechanisms including recurrent scattering, two and three-particle correlation considering short-range order shown in the diagrammatic expansion in Fig.\ref{fig1c}. The first term of second order corrections $\mathbf{\Sigma}^{(2,a)}(\omega,\mathbf{p})$  describes correlation-induced dependent scattering between pairs of particles. The other two diagrams in second order of $n_0$ give the leading terms of recurrent scattering (RS) mechanism. The RS mechanism classifies those multiple scattering paths that visit the same particle more than once \cite{lagendijk1996resonant,Lagendijk1997PRL}. This mechanism has been experimentally shown to be substantial previously in a strongly scattering medium for ultrasonic waves with $kl_s\sim 1$ \cite{Aubry2014PRL}. The two-particle RS mechanism can be separated into two parts $\mathbf{\Sigma}^{(2,b)}$  and $\mathbf{\Sigma}^{(2,c)}$, according to whether the particles for incident and emergent light waves are the same one \cite{Cherroret2016}. Since in deep near field, short-range order beyond two particles becomes important \cite{Claro1991}, we thus push the present perturbative calculation forward to consider three-particle correlation, addressing near-field effects induced by high-order particle correlations. Without increasing mathematical complexity, RS paths among three different particles are neglected, which is reasonable for dilute systems $n_0/k^3\ll1$. Therefore we retain those irreducible diagrams involving only three different particles shown in Fig. \ref{fig1c} as $\mathbf{\Sigma}^{(3,a)}(\omega,\mathbf{p})$ and $\mathbf{\Sigma}^{(3,b)}(\omega,\mathbf{p})$. These formulations enable us to understand the underlying multiple scattering physics in depth. More details for analytical calculations are shown in Supplementary Material \cite{sm}.

Since Green's tenor is the building block describing electromagnetic interaction among scatterers, distinguishing near- and far-field terms in Green{'}s tensor when calculating $\mathbf{\Sigma}(\omega,\mathbf{p})$ makes it possible to determine the effects of NFI. Specifically, a purely far-field Green's tensor is defined as $\mathbf{G}_0^{ff}(\omega,\mathbf{r})=-(\mathbf{I}-\mathbf{\hat{r}}\mathbf{\hat{r}})\exp{(ik_0r)}/(4\pi r)$, in which only the transverse component is retained . Hence a purely far-field approximated $\mathbf{\Sigma}_{ff}(\omega,\mathbf{p})$ is obtained by using $\mathbf{G}_0^{ff}(\omega,\mathbf{r})$. In the following text, results using full Green’s function are denoted as the near-field case (NFC), while the results considering only far-field interaction are denoted as far-field case (FFC).

\begin{figure*}
		%\captionsetup{labelformat=simple}
	\subfloat{
		\label{fig2a}
		\includegraphics[width=0.35\linewidth]{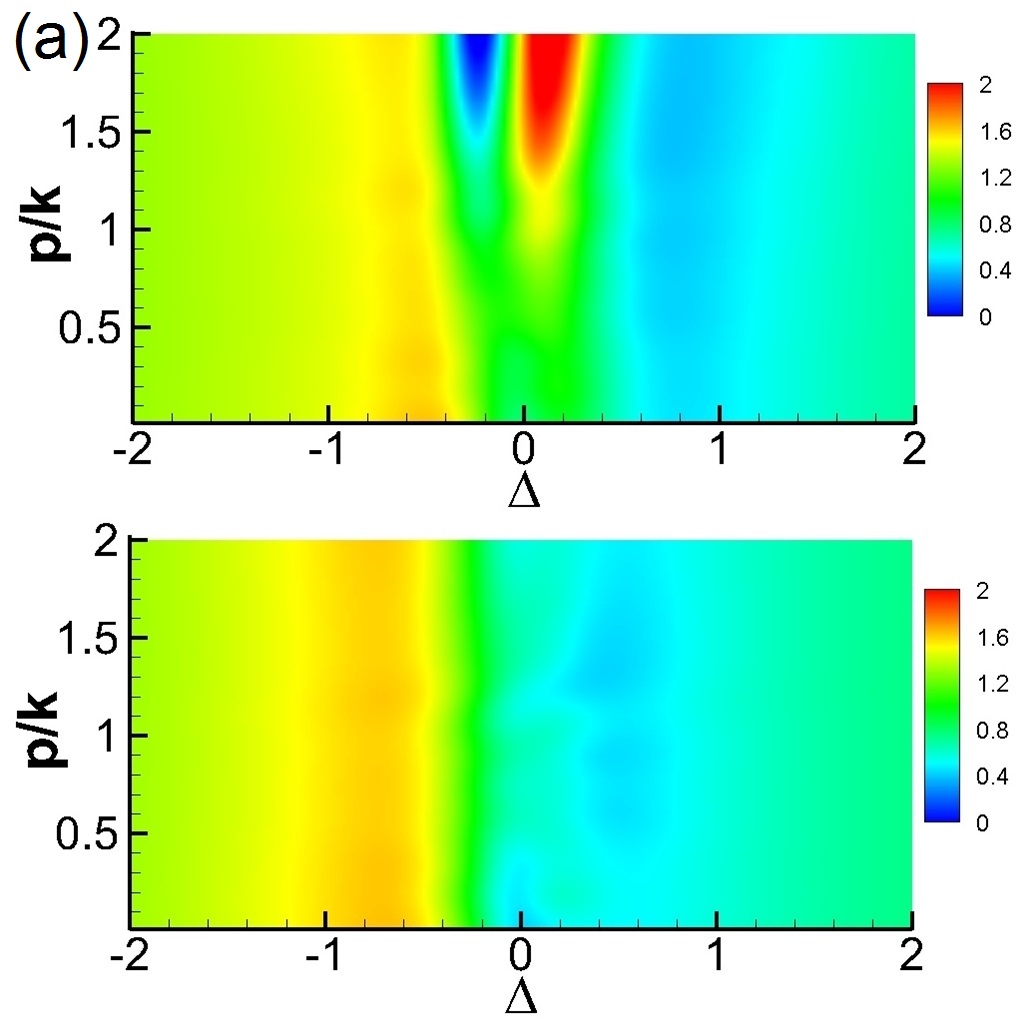}
	}
\hspace{0.01in}
	\subfloat{
		\label{fig2b}
		\includegraphics[width=0.4\linewidth]{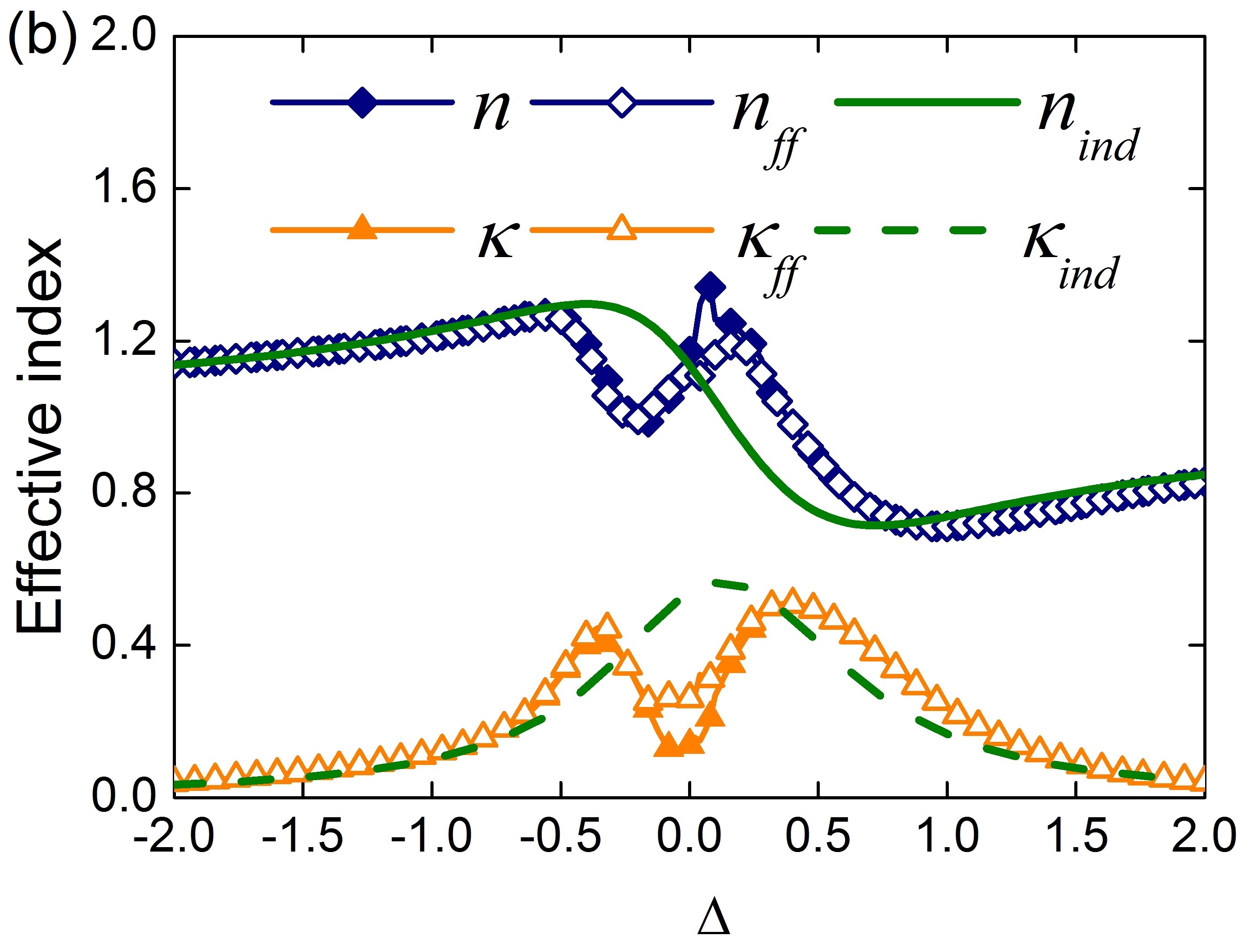}
	}
	
%	\subfloat{\label{fig2c}
%		\includegraphics[width=0.35\linewidth]{Fig2c}
%	}
	\hspace{0.01in}
	\subfloat{\label{fig2d}
		\includegraphics[width=0.4\linewidth]{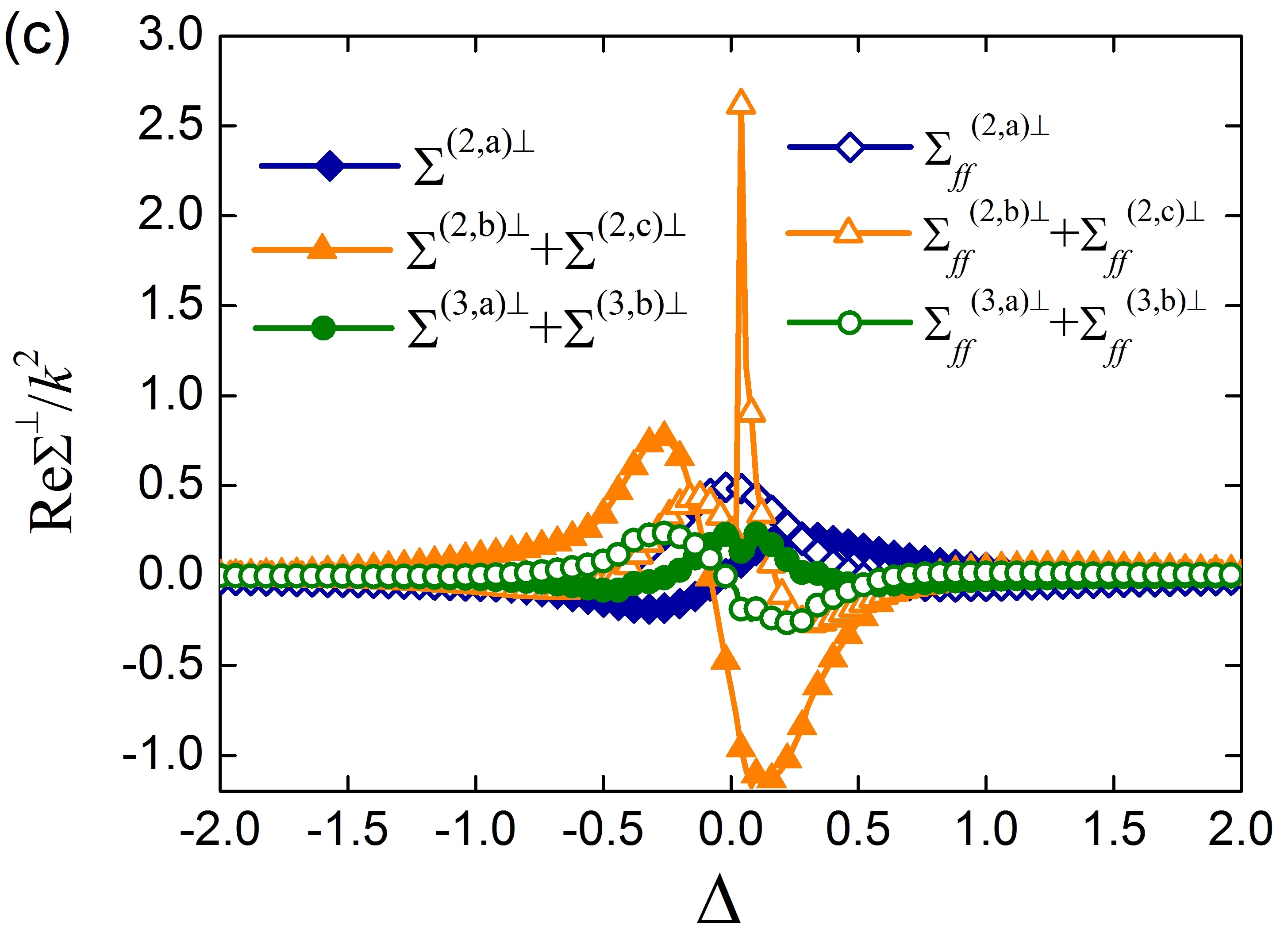}
	}
	\hspace{0.01in}
	\subfloat{\label{fig2e}
	\includegraphics[width=0.4\linewidth]{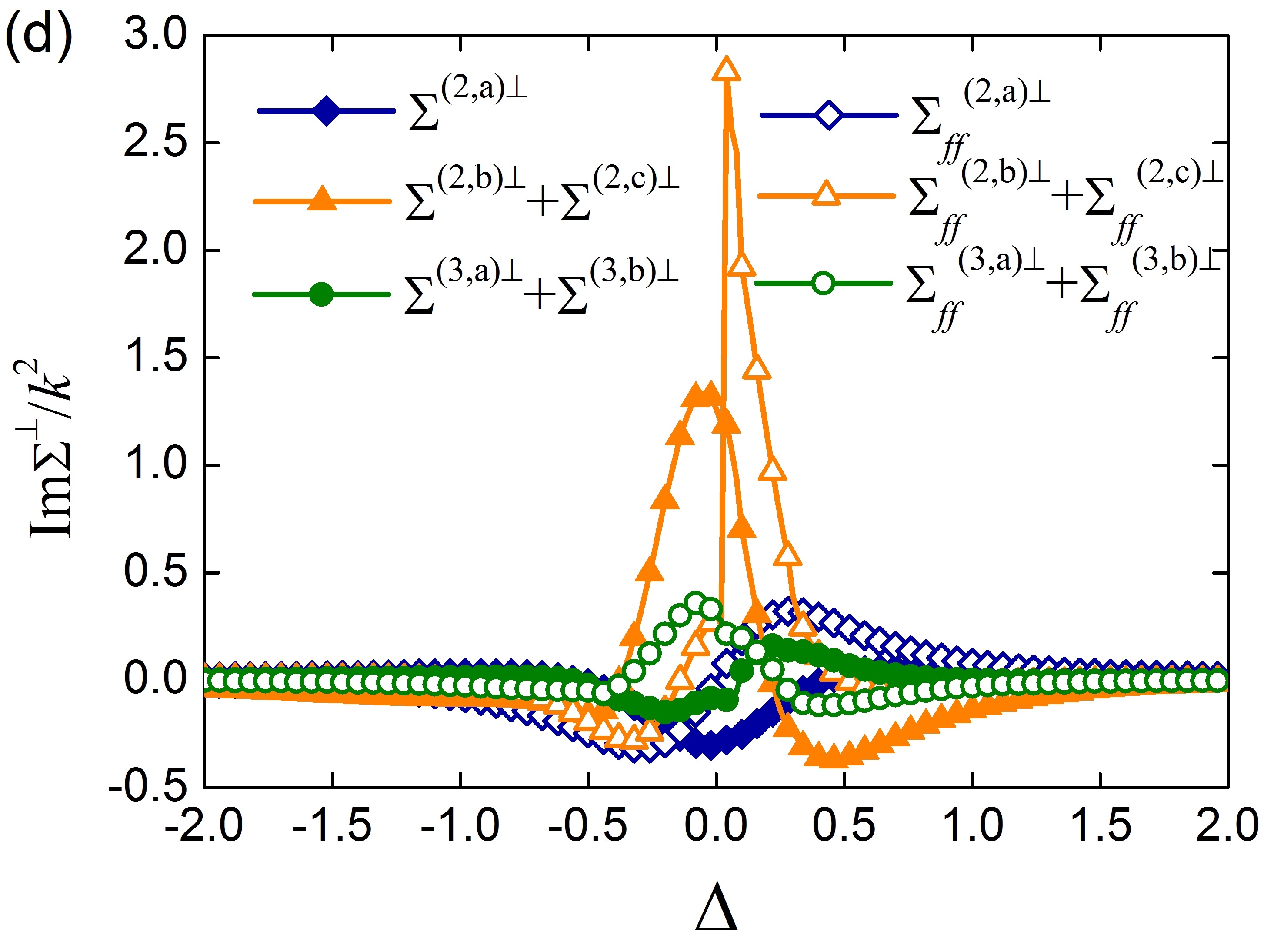}
	}
	\caption{Disordered medium containing dipole scatterers with $f_v=0.05$ and $ka=0.57$ (hard-sphere system). (a) Real part of effective permittivity for transverse waves $\mathrm{Re}\varepsilon^\bot(\omega,\mathbf{r})$ in NFC (top) and FFC (below). (b) Complex effective indices of transverse modes $n^\bot$ and $\kappa^\bot$ in NFC and FFC, compared with ISA results. (c,d) Contributions to the real (c) and imaginary (d) parts of self-energy $\Sigma^\bot$of different many-particle scattering mechanisms for the transverse mode. }
%	(c) The Ioffe-Regel parameter $\xi$. Magenta dashed line denotes $\xi=1/2$  roughly denoting a critical scattering strength for Anderson localization.
			
\end{figure*}

\textit{Hard-sphere system.}--Fig.\ref{fig2a} presents the contour in momentum and frequency spaces for the real part of analytically calculated permittivity $\mathrm{Re}\varepsilon^\bot(\omega,\mathbf{p})$ for a disordered medium with $f_v=0.05$  and $ka=0.57$ with hard-sphere potential using Percus-Yevick approximation \cite{tsang2004scattering}. In this situation, the small parameter in the diagrammatic expansion is  $n_0/k^3=0.0645\ll1$ and therefore the perturbative treatment is valid. $\mathrm{Re}\varepsilon^\bot(\omega,\mathbf{p})$  is critical for determining the strong scattering regime where scattering mean free path $l_s$ is comparable with wavelength, i.e., $\mathrm{Re}\varepsilon^\bot(\omega,\mathbf{p})=(n^\bot)^2-(\kappa^\bot)^2\le0$, leading to a photonic pseudo-bandgap. For the present hard-sphere system $\mathrm{Re}\varepsilon^\bot(\omega,\mathbf{p})\ge0$  for all possible transverse modes, while substantial differences between NFC and FFC are still observed. Real and imaginary parts of effective index $n^\bot$  and  $\kappa^\bot$  of the disordered medium are also extracted from transverse spectral function and plotted in Fig.\ref{fig2b}, where the results of FFC and ISA are also shown for comparison. In this dilute regime, the difference between dependent scattering results (including NFC and FFC) and ISA appears near the scatterer resonance when $|\Delta|<1$ , with $\Delta=(\omega-\omega_0)/\gamma $    denoting the detuning from $\omega0$. This trend is expected since single scattering cross section of each dipole is $\sigma(\Delta)=6\pi/k^2(4\Delta^2+1)$ , resulting in an effective optical size of $2\sqrt{\sigma(\Delta=1)/\pi}\sim n_0^{-1/3}$, comparable with the average particle distance. However, the effect of NFI on $n^\bot$  and  $\kappa^\bot$ is only limited near the resonance frequency $\omega_0$, indicating the particles are mainly coupled through far-field interaction (FFI) over the spectrum. Since $\kappa^\bot$ stands for extinction due to pure scattering and is related to scattering mean free path by $l_s=1/(2k\kappa^{\bot})$, giving the renormalized Ioffe-Regel parameter $\xi=n^{\bot}kl_s=n^{\bot}/2\kappa^{\bot}$ \cite{lagendijk1996resonant}. Near resonance $\Delta=0$, we find thus NFI gives a significant enhancement for $\xi$ , implying its detrimental role for Anderson localization \cite{Skipetrov2014}. To figure out how NFI affects multiple scattering processes, in Figs.\ref{fig2d} and \ref{fig2e} we show the effects from different many-particle scattering diagrams. Recurrent scattering mechanism exhibits the most striking impact and two-particle correlation and three-particle correlation effects are also substantial. Moreover, NFI drives a totally different behavior from FFI in recurrent scattering, where near resonance the former reduces $\mathrm{Re}\Sigma^\bot$, results in a higher real part of effective permittivity while the later plays the opposite role. Since for transverse propagating modes, FFI more effectively propagates on-momentum-shell, it has a stronger contribution to multiple scattering processes. On the other hand, NFI mostly propagates off-shell and decays fast, and therefore doesn{’}t contribute to multiple scattering as efficient as far-field interaction. Thus it is expected that FFI induces higher scattering for coherent waves \cite{Naraghi2016}. 

\begin{figure*}
	%\captionsetup{labelformat=simple}
	\centering
%	\subfloat{
%	\label{fig3a}
%	\includegraphics[width=0.35\linewidth]{Fig3a}
%}
	\subfloat{
	\label{fig3b}
	\includegraphics[width=0.35\linewidth]{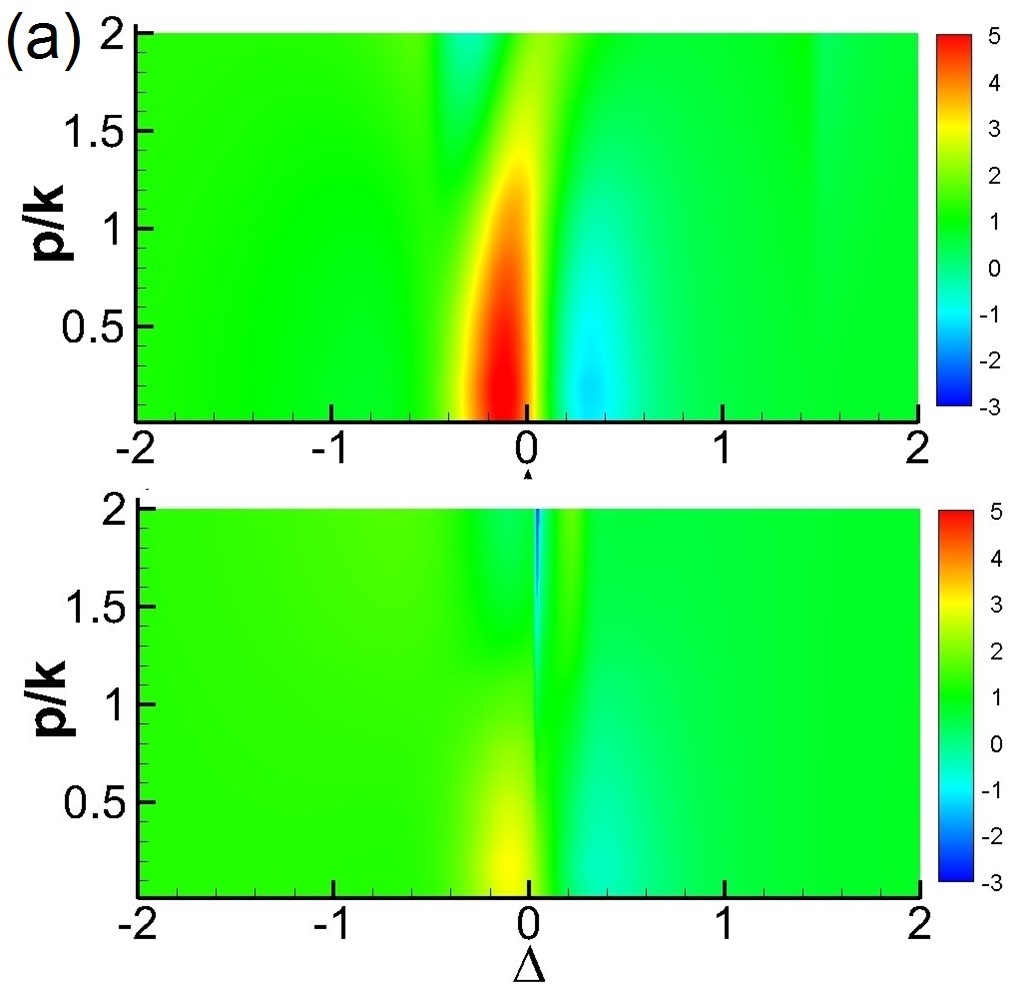}
}
\hspace{0.01in}
	\subfloat{
	\label{fig3c}
	\includegraphics[width=0.4\linewidth]{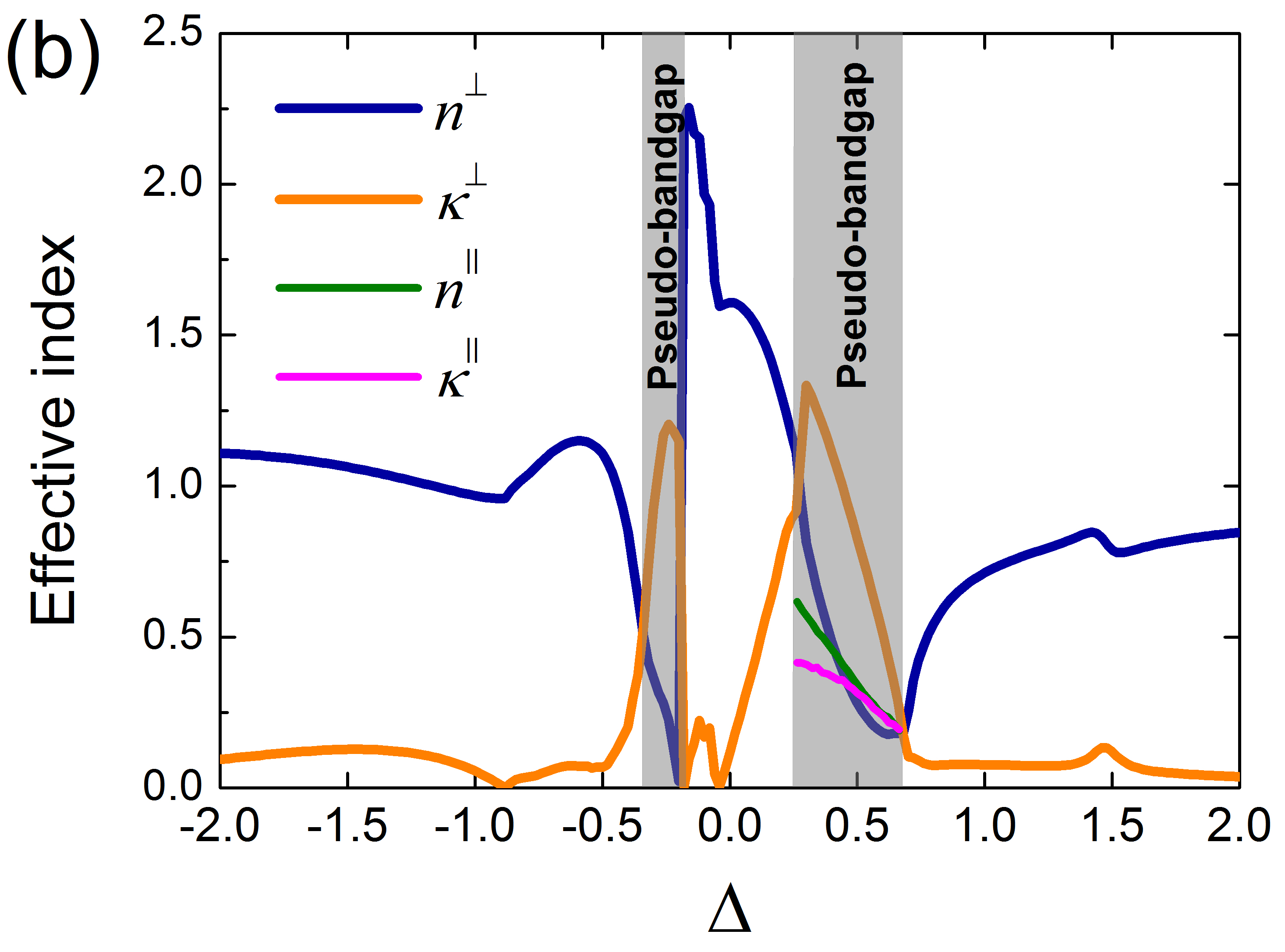}
}
\hspace{0.01in}
	\subfloat{
	\label{fig3d}
	\includegraphics[width=0.4\linewidth]{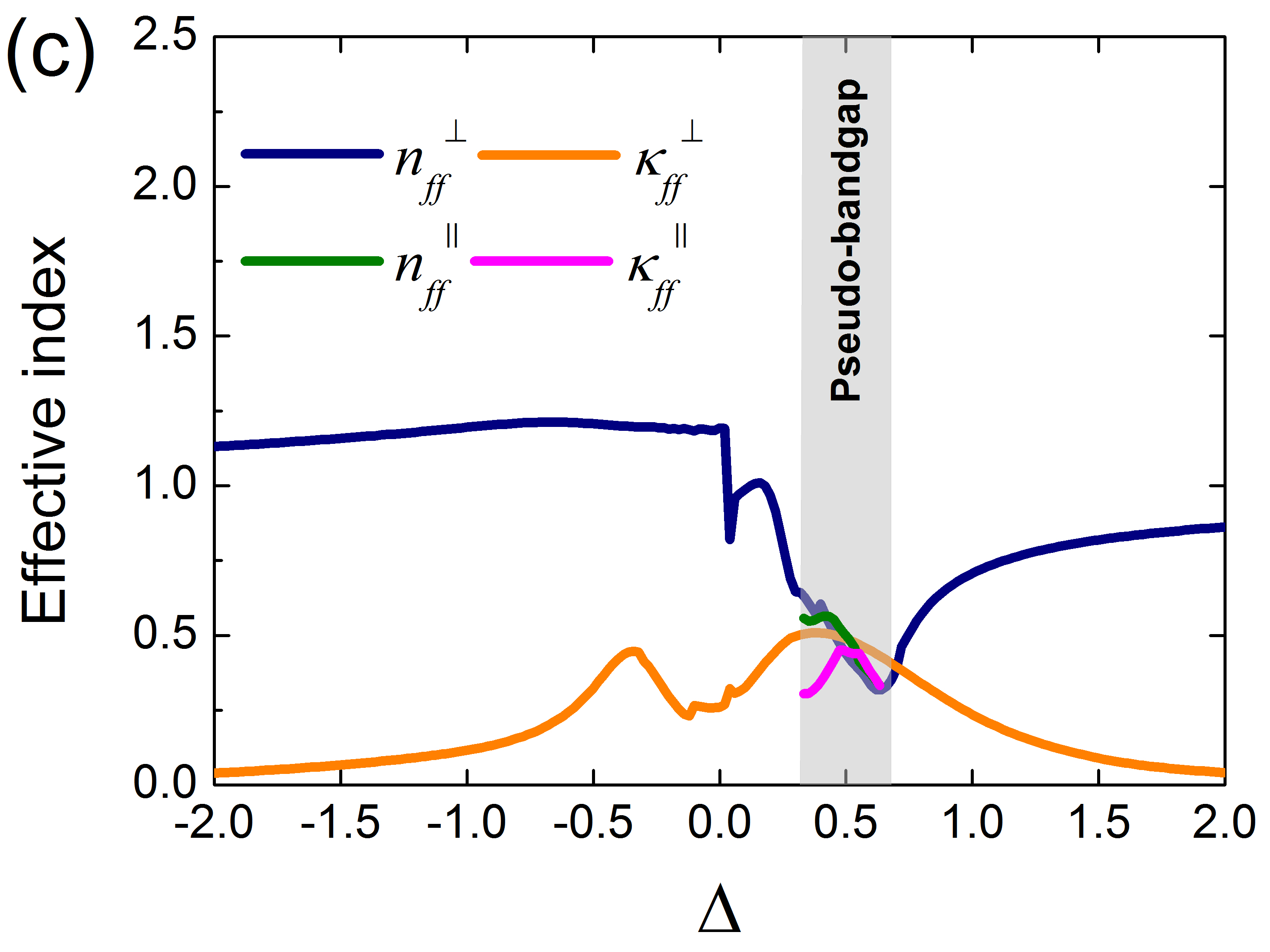}
}
	\subfloat{
	\label{fig3e}
	\includegraphics[width=0.4\linewidth]{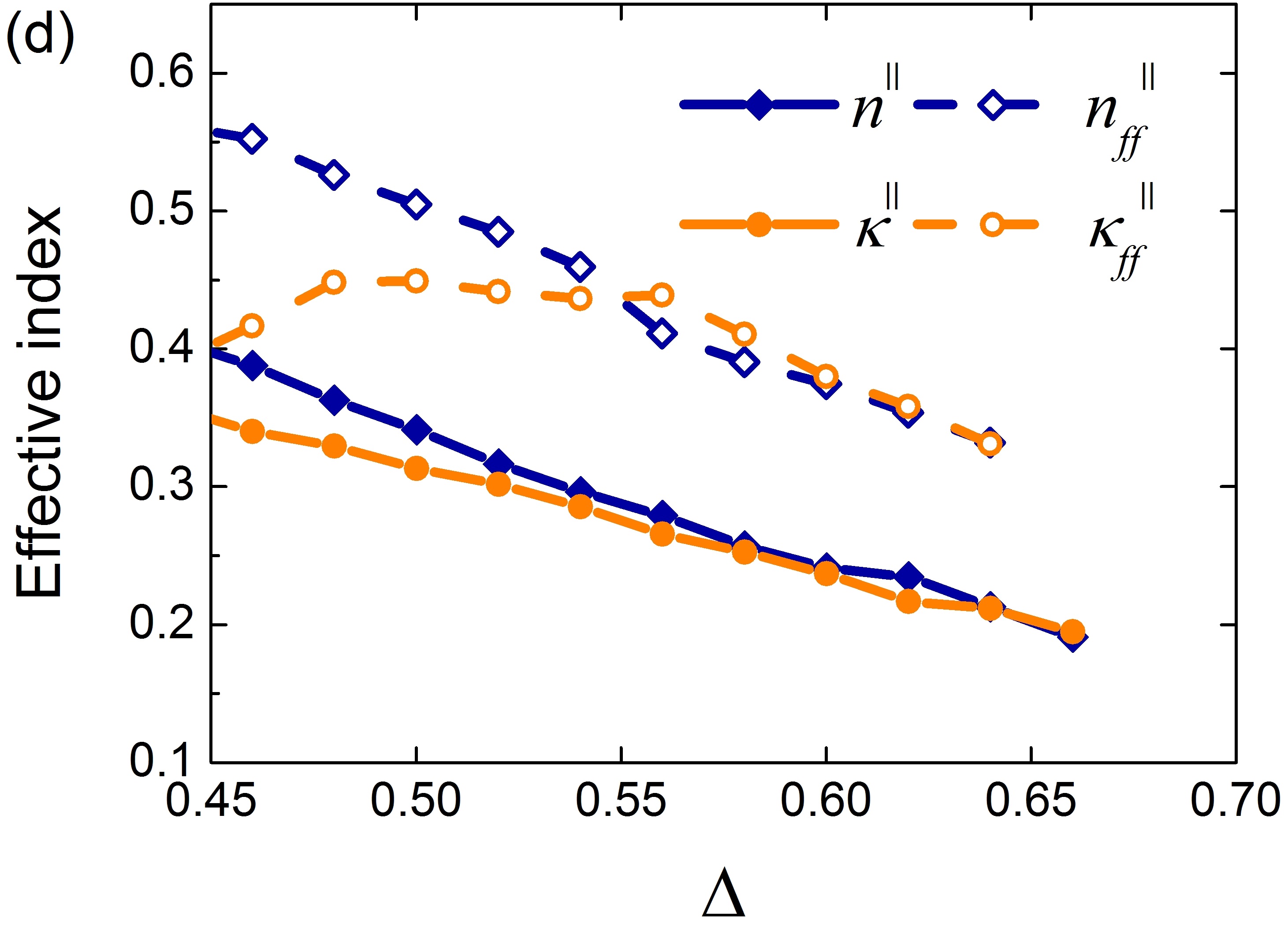}
}
	\caption{Disordered medium containing dipole scatterers with $f_v=0.05$ and $ka=0.57$ with surface adhesion $\tau=0.15$ (sticky-sphere system). (a) Real part of effective permittivity for transverse waves $\mathrm{Re}\varepsilon^\bot(\omega,p)$ in NFC (top) and FFC (below). (b,c) Complex effective indices in NFC (b) and FFC (c). The grey zones indicate pseudo-bandgap for transverse modes, where longitudinal modes are also shown. (d) Comparison of effective indices of longitudinal modes in NFC and FFC.}
\end{figure*}

\textit{Sticky sphere system.}--Short-range order and particle correlations are important factors in manipulating light flow in realistic disorder photonic media \cite{Peng2007}. To figure out the role of NFI more explicitly as well as its critical interplay with short-range order, we introduce an adhesive potential at the surfaces of particles to mimic the densely-packed disordered media instead of directly modelling the high $f_v$ behavior (infinitely short-range attraction) \cite{tsang2004scattering,Frenkel2002}. This potential brings a delta-function-like short-range order near the particle surface, quantified by the pair distribution function and thus gives an enhancement for NFI \cite{Peng2007,sm}. Fig. \ref{fig3b} presents the real part of analytically calculated transverse permittivity $\mathrm{Re}\varepsilon^\bot(\omega,p)$ for light propagation with angular frequency $\omega$ and momentum $p$. Significant negative permittivity regions are observed for both NFC and FFC, indicating these modes are strongly attenuated to be evanescent.  In addition, NFI induces more extended very-high-permittivity and very-low-permittivity zones, qualitatively indicating a stronger light-matter interaction under NFI. Effective index data retrieved from transverse spectral function is shown in Figs.\ref{fig3c} and \ref{fig3d}. For NFC in the spectral ranges of $-0.34<\Delta<0.18$ and $0.26<\Delta<0.68$, no well-defined maxima of $S^{\bot}(\omega,p)$ can be found and subsequently two pseudo-bandgaps for transverse modes are produced, which are the consequences of polaritonic behavior and all transverse modes there are evanescent \cite{lagendijk1996resonant,Schilder2016}. In FFC, only one narrower pseudo-bandgap within $0.34<\Delta<0.68$ forms. This offers a general conclusion that the present strong short-range order is able to effectively sustain and enhance multiple scattering of near fields, making particles more strongly coupled through NFI and facilitating light localization and the polaritonic behavior, as indicated by a recent experiment \cite{Silies2016}. 

%This attractive potential reads \cite{tsang2004scattering,Frenkel2002}, 
%\begin{equation}
%u(\mathbf{r})=\begin{cases}
%\infty &{0<r<s}\\
%\ln{[\frac{12\tau (b-s)}{b}]} & {s<r<b}\\
%0 &{r>b}
%\end{cases}
%\end{equation}
%where $\tau$ is the stickiness whose inverse quantifies the strength of inter-particle adhesion, $b=2a$ is particle diameter and $(b-s)$ stands for the range of potential, which is assumed to be infinitesimal because this potential is confined on particle surface.

Moreover, different with the case of hard-sphere system, pronounced local maxima of longitudinal spectral function $S^{\parallel}(\omega,p)$  also emerge for both NFC and FFC, located in the transverse pseudo-bandgap. The longitudinal mode, which exhibits a polarization vector parallel to its wave vector, is a manifestation of material degree of freedom in inhomogeneous media \cite{lagendijk1996resonant}. It is closely related to the polaritionic behavior and negative permittivity of the disordered medium where light transport is also strongly attenuated \cite{lagendijk1996resonant}. Under ISA, the present disordered medium is not dense enough to show a negative permittivity and polaritonic behavior since $\mathrm{Re}\varepsilon_{ISA}^\parallel=1-9f_v\Delta/\left[(ka)^3(4\Delta^2+1)\right]$ is always positive. Thus it indeed is the present strong short-range order that gives rise to a local high density of particles and facilitates the formation of longitudinal modes. The longitudinal mode exhibits a very anomalous dispersion curve, confirming that it undergoes a strong attenuation in this region. In Fig.\ref{fig3e} for an enlarged illustration for longitudinal modes, it is shown that the longitudinal mode in NFC possesses a smaller propagation constant $K^\parallel=n^\parallel k$ and a lower extinction coefficient $\kappa^\parallel$, indicating a longer longitudinal wavelength and propagation length. Therefore, NFI indeed enables a stronger collective excitation involving more particles. 
\begin{figure*}
	%\captionsetup{labelformat=simple}
	\centering
	\subfloat{
		\label{fig4a}
		\includegraphics[width=0.4\linewidth]{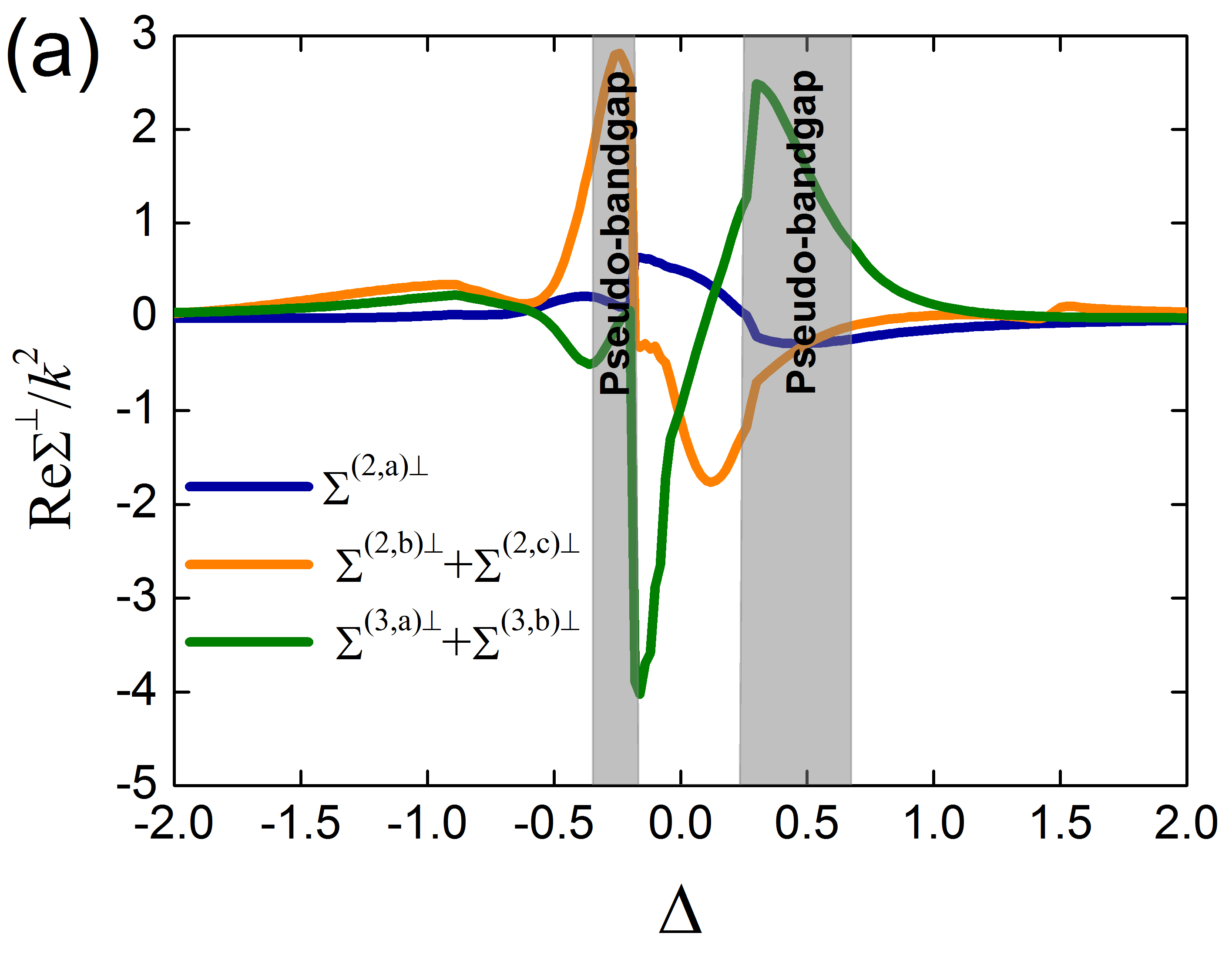}
	}
	\hspace{0.01in}
	\subfloat{
		\label{fig4b}
		\includegraphics[width=0.4\linewidth]{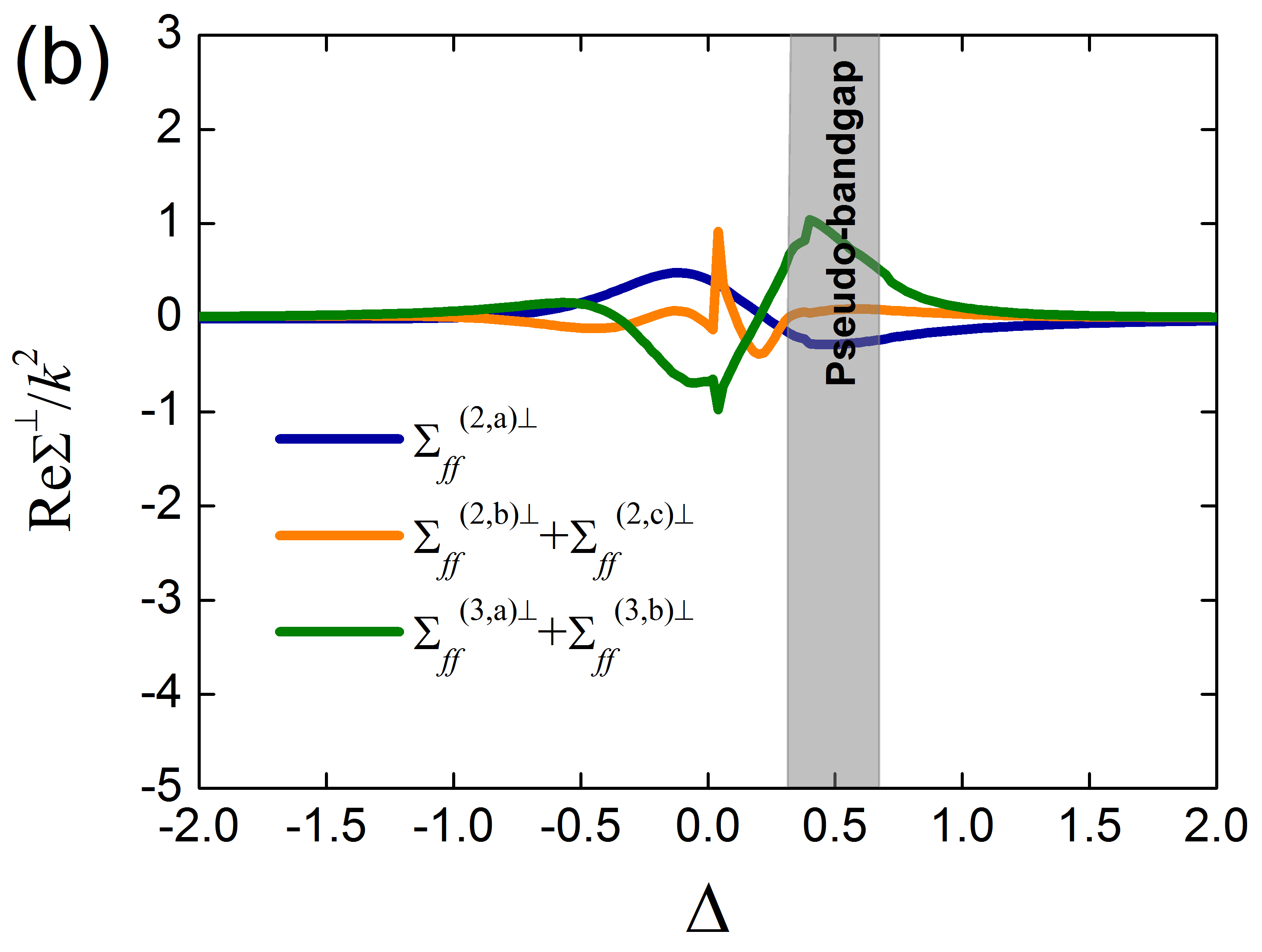}
	}
	\hspace{0.01in}
	\subfloat{
		\label{fig4c}
		\includegraphics[width=0.4\linewidth]{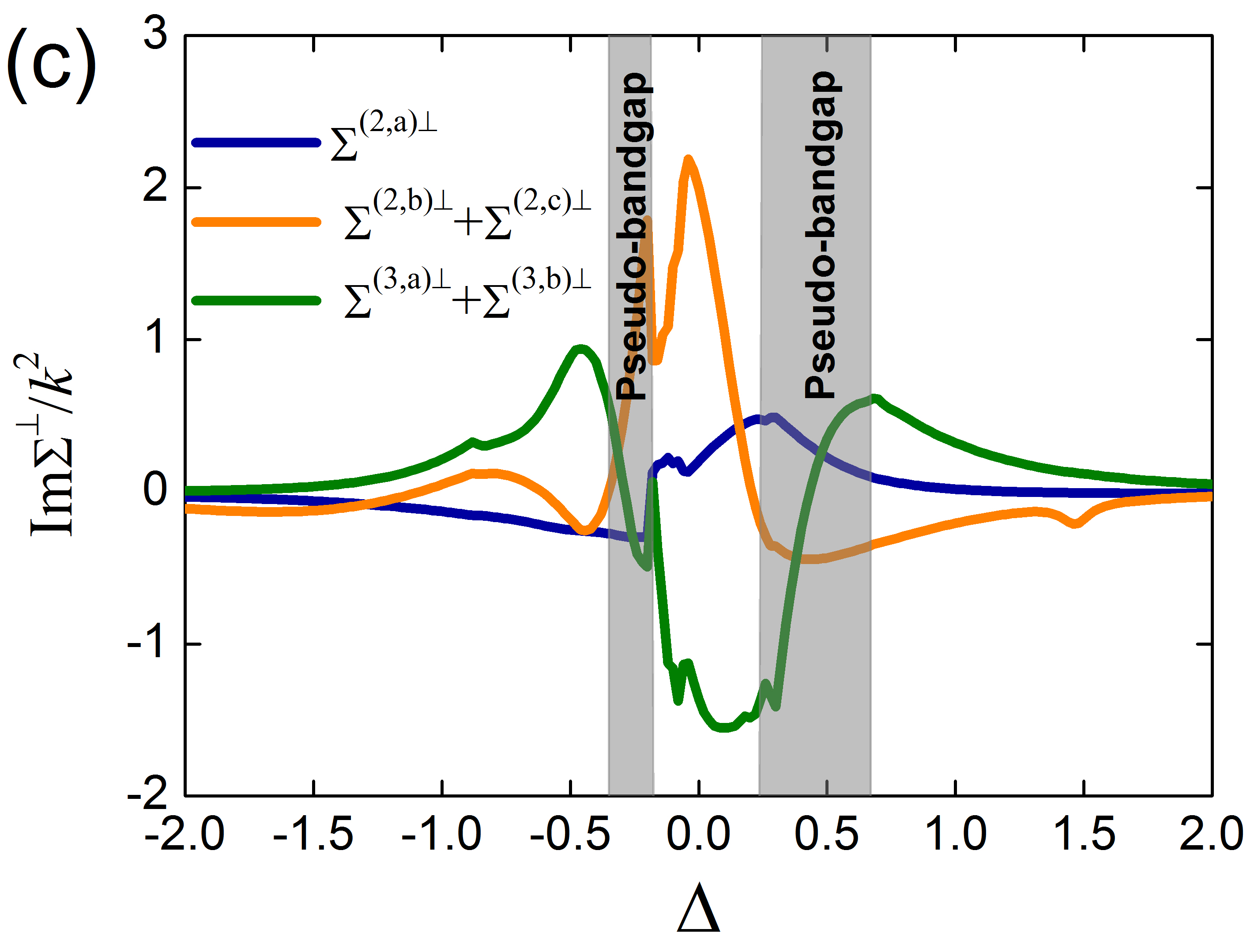}
	}
	\subfloat{
		\label{fig4d}
		\includegraphics[width=0.4\linewidth]{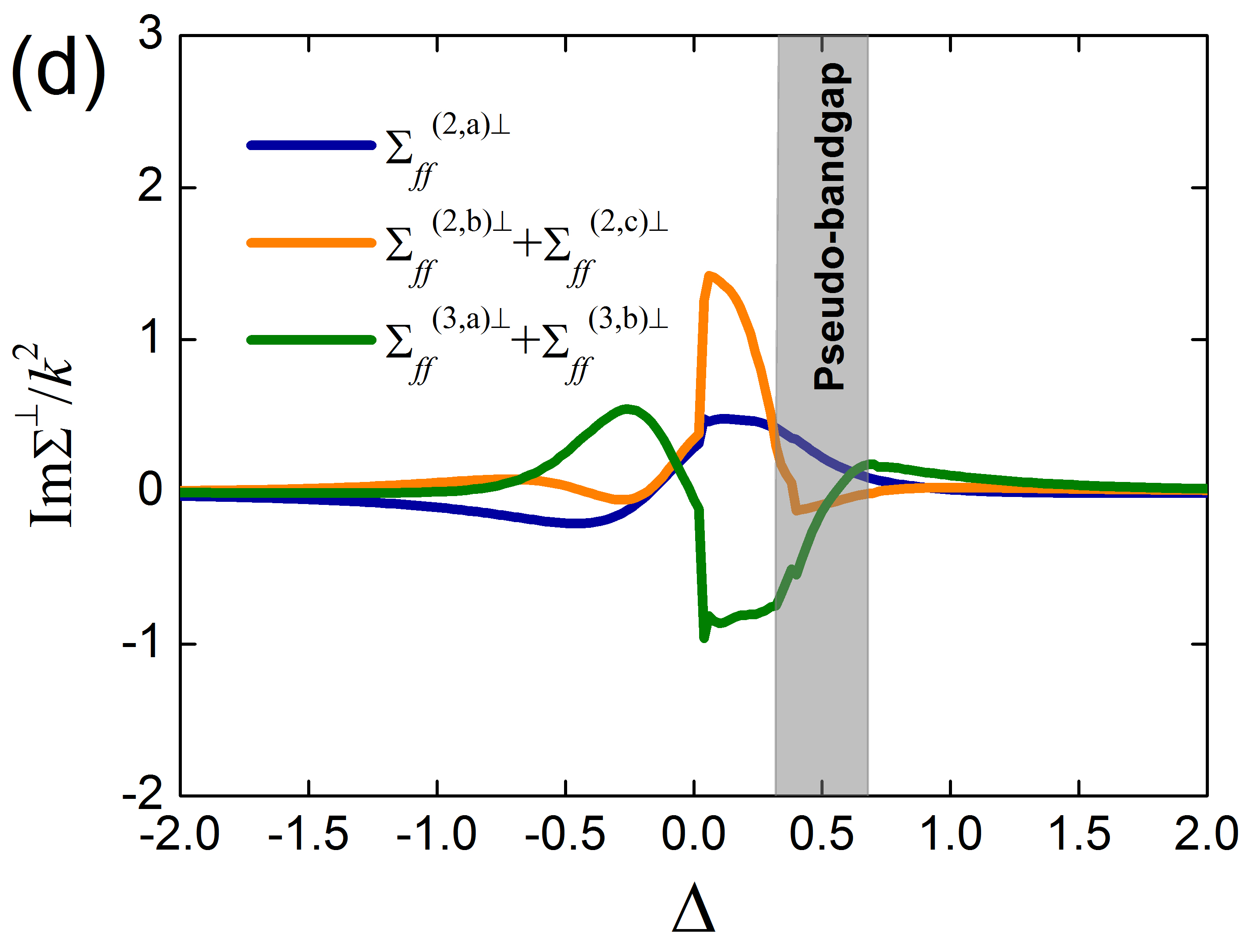}
	}
	%\subfloat{
	%	\label{fig4e}
	%	\includegraphics[width=0.4\linewidth]{Fig4e}
	%}
	\caption{Different many-particle mechanisms in the sticky-sphere system. (a,c) Contributions of many-particle mechanisms to real parts (a) and imaginary parts (c) of self-energy for transverse modes in NFC. (b,d) Contributions of many-particle mechanisms to real parts (b) and imaginary parts (d) of self-energy of transverse modes in FFC.}\label{fig4}
\end{figure*}
In the spectra below the pseudo-bandgap, a significant difference is also observed between NFC and FFC over the spectral range of $-2<\Delta<-0.34$ when particles are rather weakly scattering. In Fig.\ref{fig4} depicting the contributions of different multiple scattering mechanisms, it is observed that NFI indeed has a much larger contribution to recurrent scattering and three-particle correlation when compared with Fig.\ref{fig2d} in the hard-sphere system. Particularly, in the spectral range of $-2<\Delta<-1.16$, NFI results in a higher scattering strength and a smaller index of refraction. According to Fig.\ref{fig4}, this is because NFI more effectively couples particles together through recurrent scattering and three-particle scattering mechanisms, indicating a stronger light-matter interaction and a longer propagation wavelength, while FFI has little impact on this partial spectral profile due to the weak coupling between weakly-scattering single particles. Moreover, in the spectral range of $-1.16<\Delta<-0.34$ , NFI still leads to a lower index of refraction yet supressing scattering strength when compared to FFI, which is attributed to the fact that three-particle correlation mechanism strongly reduces scattering in this region shown in Fig.\ref{fig4c}. In this regime, the strong NFI among particles are transformed from evanescent waves into propagating waves, resulting in a reduction of scattering and perhaps making new transport channels open \cite{Naraghi2015}. On the other hand, near single particle resonance ($\Delta\sim0$), NFI causes a giant increase refractive index and a dip in scattering strength (Figs.\ref{fig3c} and \ref{fig3d}). The former mainly derives from the NFI-enhanced three-particle correlation while the latter is because NFI-enhanced recurrent scattering strongly reduces scattering.  In a word, the introduction of the present strong short-range order indeed enhances the contribution of NFI, making it more effectively coupled to multiple scattering trajectories. However, whether NFI undermines or enhances scattering and localization relies heavily on the interplay among frequency-dependent single particle scattering properties, short-range order and NFI simultaneously.

\textit{Local density of states (LDOS).}--In modern optics, local density of states (LDOS) is a central concept as a characterization of density of modes per frequency at certain position \cite{lagendijk1996resonant,narayanaswamy2010dyadic,leseurEPJST2017}. It is directly linked with the criterion of Anderson localization \cite{lagendijk1996resonant} and spontaneous decay rate of quantum emitters coupled to disordered media \cite{Wiersma2008}. In addition, its statistics and correlations also reflect the microscopic information of random media \cite{Sapienza2011,DeSousa2016}. The configurational averaged LDOS can be derived from averaged Green's tensor for an unbounded disordered medium as \cite{lagendijk1996resonant,sheng2006introduction,narayanaswamy2010dyadic}
\begin{equation}
\rho_c(\omega,\mathbf{R}=0)=\frac{n^{\bot}\omega^2}{\pi^2 c^3}-\frac{1}{2\pi^3\omega}\mathrm{Im}\int{\frac{p^2dp}{\varepsilon^{\parallel}(\omega,p)}}
\label{ldos}
\end{equation}
which contains both transverse ($\rho_c^\bot$) and longitudinal ($\rho_c^\parallel$ ) components. For the present non-absorbing medium, the first part denotes the contribution of coherent waves, while the second part indicates incoherent waves which, actually, give rise to diffusive decay channels, unlike the role of nonradiative decay channels in absorptive media. Actually, longitudinal part of LDOS, attributed to longitudinal modes, is related to the imaginary part of longitudinal permittivity $\mathrm{Im}\varepsilon^\parallel(\omega,p)$. In non-absorbing random media, it is then treated as the loss purely due to random scattering, connected with the fluctuating part of the scattering field \cite{Froufe-Perez2007}. The decayed energy associated with longitudinal LDOS is then scattered and becomes diffusive. Since the integral for $\rho_c^\parallel$ in Eq.(\ref{ldos}) diverges, an upper-limit cut-off of longitudinal momentum is necessary. This cut-off actually involves realistic local structural details of the coupled quantum emitter, making analytical determination of the non-radiative decay term unclear \cite{leseurEPJST2017}. For a given emitter/random-media configuration, this cut-off of momentum $p_{max}$  is directly dependent on the size of cavity circumscribing the emitter as $R_0^{-1}$ formed by random media, namely, the local field effect \cite{Barnett1996,LagendijkPRL1998a,LagendijkPRL1998b,DeSousa2016,leseurEPJST2017}. Thus the underlying physics is trivial in essence and readily described by Eq.(\ref{ldos}). To avoid direct evaluation of Eq.\ref{ldos}, here we present an exact numerical calculation on LDOS based on the coupled dipole approach \cite{Caze2013,sm} for the spherical volume $R=10a$, where the probing dipole emitter is positioned at the center of this volume and particles are restricted not to overlap with it. Such an emitter acts as interstitial that does not break the correlations between the scatterers \cite{LagendijkPRL1998b}. For generating sticky particles, we implemented Kranendonk-Frenk algorithm \cite{Frenkel2002}.

% Note LDOS entering the Einstein spontaneous emission coefficient implies only the propagating LDOS $\rho_c^\bot$, and it is also $\rho_c^\bot$ that acts as an indicator for Anderson localization \cite{lagendijk1996resonant}. Therefore, the propagating part of LDOS is simply proportional with the real part of transverse effective index $n^\bot$ if the transverse modes are well-defined\cite{Barnett1996}, and above discussions on $n^\bot$ also apply to $\rho_c^\bot$.
\begin{figure}
	%\captionsetup{labelformat=simple}
	\centering
	\includegraphics[width=0.8\linewidth]{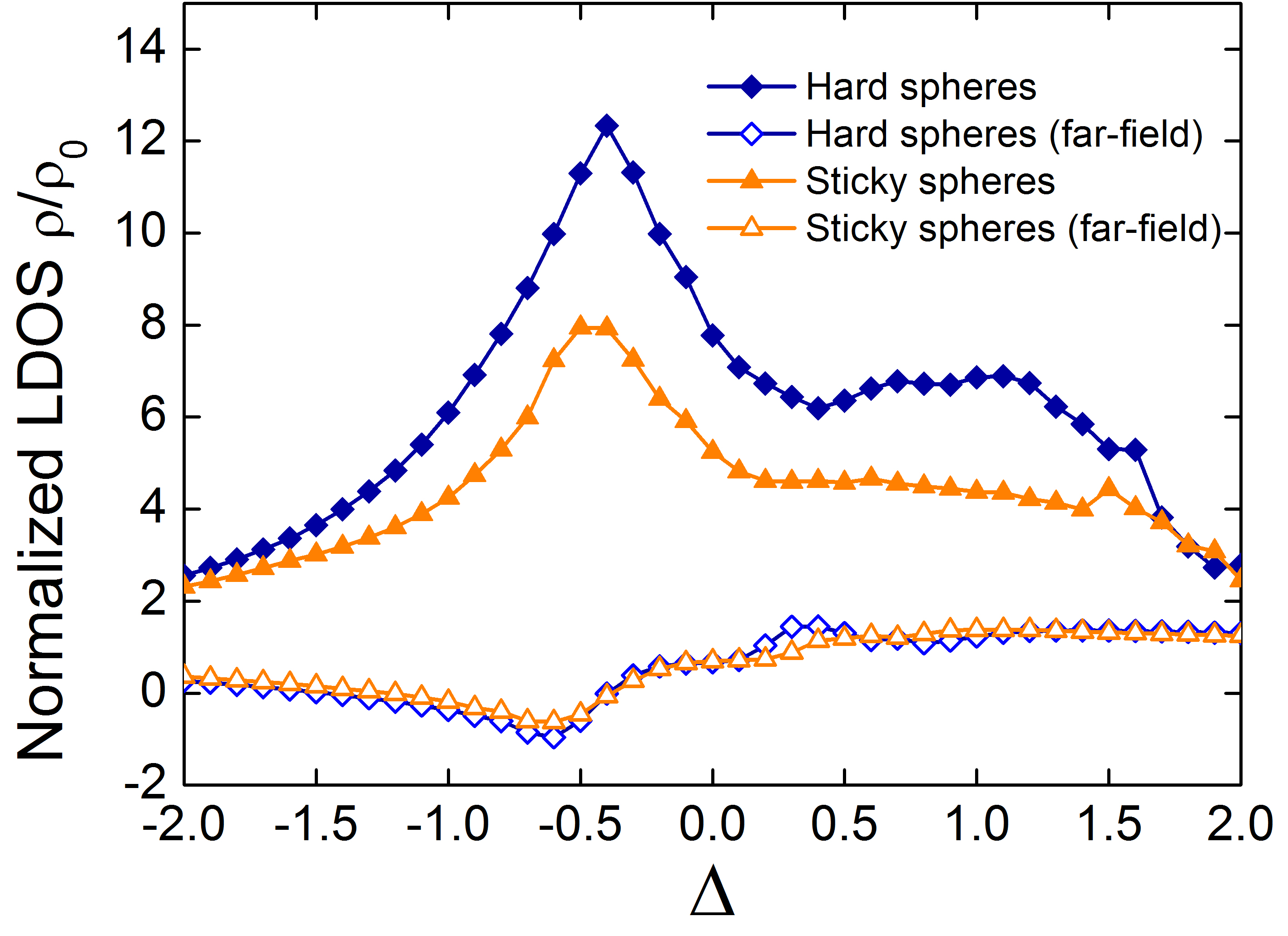}
	\label{fig5}
	\caption{Numerically calculated configurational averaged LDOS $\rho_c$ normalized by free space LDOS $\rho_0$.}\label{LDOS}
\end{figure}

The calculated LDOS is shown in Fig.\ref{LDOS}. A large difference between the hard-sphere and sticky-sphere systems considering NFI is observed, which addresses the failure of typical methods of ISA and even the mean-field Lorentz-Lorenz formula \cite{Schilder2016} both neglecting the short-range order in particle correlations, while no difference is recognized for the two systems under only FFI. This leads to us a general conclusion that NFI can enhance LDOS. The results are well understood based on our theoretical results. In the hard-sphere system, the maximum of LDOS is located at around $\Delta\sim-0.5$. This is where both $n^\bot$ and $\kappa^\bot$ reach relatively large values, which correspond to coherent and incoherent decay channels respectively. However, for the FFC, this argument alone is not able to explain the deviation simply by using $n^\bot$ and $\kappa^\bot$ because at $\Delta\sim-0.5$ the effective indices are also close to NFC. This, on the other hand, is due to the point that LDOS is more relevant with the longitudinal modes and local environment of the emitter. Since formation of a local cavity by the disordered media for the emitter is a many-particle process \cite{andreasen2011modes}, the high-degree of on-shell propagation of FFI among particles leads to a strong leakage of modes. Thus LDOS is greatly reduced around the emitter. It is noted that neglecting NFI leads to the unphysical reduction of LDOS to negative values. In the meanwhile, NFI between multiple particles, can still sustain more incoherent longitudinal modes as well as a smaller effective cavity and thus LDOS. For the sticky-sphere system, even in the pseudo-bandgap, LDOS never approaches zero due to the existence of strong longitudinal, incoherent decaying channels therein. 

\textit{Summary.}--To summarize, we have formally analyzed the role of NFI on light transport in disordered media. Formulations on many-particle scattering including recurrent scattering, three particle correlations are analytically derived. We find NFI leads to a stronger collective excitation involving more particles. More weakly-decayed longitudinal modes are also formed through NFI. We also demonstrate that strong short-range order enables one to enhance NFI and thus tune local density of states (LDOS). We expect present discussions to be fruitful for understanding coherent light scattering physics and inspiring applications in disordered media. We also anticipate profound implications can be offered on the cooperative effects in optically dense cold atom systems showing a great potential for applications in quantum optics and quantum information science \cite{guerin2017light,Kupriyanov20171}.

\begin{acknowledgments}
We thank the financial support from the National Natural Science Foundation of China (Nos. 51636004, 51476097), Shanghai Key Fundamental Research Grant (Nos. 18JC1413300, 16JC1403200), and the Foundation for Innovative Research Groups of the National Natural Science Foundation of China (No.51521004).
\end{acknowledgments}
\bibliography{nfi2018}

\end{document}

\begin{center}
	\textbf{\Large{Supplemental Material for}}
	
	\textbf{\Large{"Role of Near-Field Interaction on Light Transport in Disordered Media"}}
\end{center}
	
\vskip 5mm

\section{Derivation of $\bm{\Sigma}^{(2,a)}$, $\bm{\Sigma}^{(2,b)}$ and $\bm{\Sigma}^{(2,c)}$}
The free-space Green's function is expressed as follows
\begin{equation}
\begin{split}
\mathbf{G}_{0}(\omega,\mathbf{r})=&-(\frac{i}{kr}-\frac{1}{(kr)^2}+1)\frac{\exp{(ikr)}}{4\pi r}(\mathbf{I}-\mathbf{\hat{r}}\mathbf{\hat{r}})\\
&+2(\frac{i}{kr}-\frac{1}{(kr)^2})\frac{\exp{(ikr)}}{4\pi r}\mathbf{\hat{r}}\mathbf{\hat{r}}+\frac{\delta(\mathbf{r})}{3k^{2}}\mathbf{I}\\
&=G_0^{\bot}(r)(\mathbf{I}-\mathbf{\hat{r}}\mathbf{\hat{r}})+G_0^{\parallel}(r)\mathbf{\hat{r}}\mathbf{\hat{r}}+\frac{\delta(\mathbf{r})}{3k^{2}}\mathbf{I}
\end{split}
\end{equation}
where we discriminate transverse part and longitudinal part, and let $\mathbf{P}(\mathbf{r})=\mathbf{I}-\mathbf{\hat{r}}\mathbf{\hat{r}}$ and $\mathbf{Q}(\mathbf{r})=\mathbf{\hat{r}}\mathbf{\hat{r}}$ to be the transverse and longitudinal projection tensors respectively.
Thus the diagrammatic terms are given by
\begin{equation}
\begin{split}
\mathbf{\Sigma}^{(2,a)}(\omega,\mathbf{p})=&n_0^2t^2\int{d^3\mathbf{r}\mathbf{G}_0(\mathbf{r})h_2(\mathbf{r})\exp{(i\mathbf{p}\cdot\mathbf{r})}}=-\frac{n_0^2t^2}{3k^2}\mathbf{I}\\
&+4\pi n_0^2t^2\int{r^2drh_2(r)\left[G_0^{\bot}(r)j_0(pr)\mathbf{I}+\left(G_0^{\parallel}(r)-G_0^{\bot}(r)\right)
	\left(\begin{matrix}
	&\frac{j_1(pr)}{pr} &  & \\
	& &\frac{j_1(pr)}{pr} &  \\
	&  & &\frac{j_0(pr)pr-j_1(pr)}{pr}
	\end{matrix}\right)
	\right]}
\end{split}
\end{equation}

\begin{equation}
\begin{split}
\mathbf{\Sigma}^{(2,b)}(\omega,\mathbf{p})=&n_0^2t^3\int{d^3\mathbf{r}\frac{\mathbf{G}_0^2(\mathbf{r})[1+h_2(\mathbf{r})]}{\mathbf{I}-t^2\mathbf{G}_0^2(\mathbf{r})}}\\=&4\pi n_0^2t^3\mathbf{I}\int{r^2dr\left[\frac{2}{3}\frac{G_0^{\bot 2}(r)}{1-t^2G_0^{\bot 2}(r)}+\frac{1}{3}\frac{G_0^{\parallel 2}(r)}{1-t^2G_0^{\parallel 2}(r)}\right]\left[1+h_2(r)\right]}
\end{split}
\end{equation}
\begin{equation}
\begin{split}
\mathbf{\Sigma}^{(2,c)}(\omega,\mathbf{p})=&n_0^2t^4\int{d^3\mathbf{r}\frac{\mathbf{G}_0^3(\mathbf{r})[1+h_2(\mathbf{r})]}{\mathbf{I}-t^2\mathbf{G}_0^2(\mathbf{r})}\exp{(i\mathbf{p}\cdot\mathbf{r})}}=4\pi n_0^2t^4\int{r^2dr[1+h_2(r)]}\\
&\times \left\{\frac{G_0^{\bot 3}(r)}{1-t^2G_0^{\bot 2}(r)}j_0(pr)\mathbf{I}+\left[\frac{G_0^{\parallel 3}(r)}{1-t^2G_0^{\parallel 2}(r)}-\frac{G_0^{\bot 3}(r)}{1-t^2G_0^{\bot 2}(r)}\right]
\left(\begin{matrix}
&\frac{j_1(pr)}{pr} &  & \\
& &\frac{j_1(pr)}{pr} &  \\
&  & &\frac{j_0(pr)pr-2j_1(pr)}{pr}
\end{matrix}\right)\right\}
\end{split}
\end{equation}

\section{Derivation of $\bm{\Sigma}^{(3)}$}
\begin{equation}
\mathbf{\Sigma}(\omega,\mathbf{p})=\mathbf{\Sigma}^{(3,a)}(\omega,\mathbf{p})+\mathbf{\Sigma}^{(3,b)}(\omega,\mathbf{p})
\end{equation}
\begin{equation}
\mathbf{\Sigma}^{(3,a)}(\omega,\mathbf{p})=n_0^3t^3\int{d^3\mathbf{r}d^3\mathbf{r}'\mathbf{G}_0(\mathbf{r}-\mathbf{r}')\mathbf{G}_0(\mathbf{r}')h_2(\mathbf{r})\exp{(i\mathbf{p}\cdot\mathbf{r})}}
\end{equation}
\begin{equation}
\mathbf{\Sigma}^{(3,b)}(\omega,\mathbf{p})=n_0^3t^3\int{d^3\mathbf{r}d^3\mathbf{r}'\mathbf{G}_0(\mathbf{r}-\mathbf{r}')\mathbf{G}_0(\mathbf{r}')h_3(\mathbf{r},\mathbf{r}')\exp{(i\mathbf{p}\cdot\mathbf{r})}}
\end{equation}
The subscript 3 in $h_3(\mathbf{r},\mathbf{r}')$  denotes the simultaneous three particle correlation. To calculate Eq.8 analytically, Kirkwood superposition approximation (KSA) for three-particle correlation is used, which is valid only for dilute systems \cite{Kirkwood1936}. 
\begin{equation}
\begin{split}
h_3(\mathbf{r},\mathbf{r}')=&h_2(\mathbf{r}-\mathbf{r}')h_2(\mathbf{r}')+h_2(\mathbf{r})h_2(\mathbf{r}')\\
&+h_2(\mathbf{r}-\mathbf{r}')h_2(\mathbf{r})+h_2(\mathbf{r}-\mathbf{r}')h_2(\mathbf{r}')h_2(\mathbf{r})
\end{split}
\end{equation}
\begin{equation}
\begin{split}
\mathbf{\Sigma}^{(3)}(\omega,\mathbf{p})=&\mathbf{\Sigma}^{(3,a)}+\mathbf{\Sigma}^{(3,b)}=n_0^3t^3\int{d^3\mathbf{r}d^3\mathbf{r}'\left[G_0^{\bot}(\mathbf{r}-\mathbf{r}')\mathbf{P}(\mathbf{r}-\mathbf{r}')+G_0^{\parallel}(\mathbf{r}-\mathbf{r}')\mathbf{Q}(\mathbf{r}-\mathbf{r}')+\frac{\delta(\mathbf{r}-\mathbf{r}')}{3k^2}\mathbf{I}\right]}\\
&\times \left[G_0^{\bot}(\mathbf{r}')\mathbf{P}(\mathbf{r}')+G_0^{\parallel}(\mathbf{r}')\mathbf{Q}(\mathbf{r}')+\frac{\delta(\mathbf{r}')}{3k^2}\mathbf{I}\right]\exp{(i\mathbf{p}\cdot\mathbf{r})}\\
&\times \left[h_2(\mathbf{r})+h_2(\mathbf{r}-\mathbf{r}')h_2(\mathbf{r}')+h_2(\mathbf{r})h_2(\mathbf{r}')+h_2(\mathbf{r}-\mathbf{r}')h_2(\mathbf{r})+h_2(\mathbf{r}-\mathbf{r}')h_2(\mathbf{r}')h_2(\mathbf{r})\right]\\
=&n_0^3t^3\int{d^3\mathbf{r}d^3\mathbf{r}'\left[G_0^{\bot}(\mathbf{r}-\mathbf{r}')\mathbf{P}(\mathbf{r}-\mathbf{r}')+G_0^{\parallel}(\mathbf{r}-\mathbf{r}')\mathbf{Q}(\mathbf{r}-\mathbf{r}')\right]}\times \left[G_0^{\bot}(\mathbf{r}')\mathbf{P}(\mathbf{r}')+G_0^{\parallel}(\mathbf{r}')\mathbf{Q}(\mathbf{r}')\right]\\
&\times\exp{(i\mathbf{p}\cdot\mathbf{r})}\left[h_2(\mathbf{r})+h_2(\mathbf{r}-\mathbf{r}')h_2(\mathbf{r}')+h_2(\mathbf{r})h_2(\mathbf{r}')+h_2(\mathbf{r}-\mathbf{r}')h_2(\mathbf{r})+h_2(\mathbf{r}-\mathbf{r}')h_2(\mathbf{r}')h_2(\mathbf{r})\right]\\
&-\frac{2 n_0^3t^3}{3k^2}\int d^3\mathbf{r}\mathbf{G}_0(\mathbf{r})h_2(\mathbf{r}) 
\end{split}
\end{equation}
To solve above integral, we change the variables by $\mathbf{r}_0$
\begin{equation}
\begin{split}
\mathbf{\Sigma}^{(3)}(\omega,\mathbf{p})=&n_0^3t^3\int{d^3\mathbf{r_0}d^3\mathbf{r}'\left[G_0^{\bot}(\mathbf{r}_0)\mathbf{P}(\mathbf{r}_0)+G_0^{\parallel}(\mathbf{r}_0)\mathbf{Q}(\mathbf{r}_0)\right]}\times \left[G_0^{\bot}(\mathbf{r}')\mathbf{P}(\mathbf{r}')+G_0^{\parallel}(\mathbf{r}')\mathbf{Q}(\mathbf{r}')\right]\\
&\times\exp{\left[i\mathbf{p}\left(\mathbf{r}_0+\mathbf{r}'\right)\right]}\left\{h_2(\mathbf{r}_0)h_2(\mathbf{r}')+h_2(\mathbf{r}_0+\mathbf{r}')\left[1+h_2(\mathbf{r}')\right]\left[1+h_2(\mathbf{r}_0)\right]\right\}\\
&-\frac{2 n_0^3t^3}{3k^2}\int d^3\mathbf{r}\mathbf{G}_0(\mathbf{r})h_2(\mathbf{r}) 
\end{split}
\end{equation}
The principal part (PS) containing integrals of principal parts of Green{'}s tensor is denoted by
\begin{equation}
\begin{split}
\mathbf{\Sigma}^{(3,PS)}(\omega,\mathbf{p})=&n_0^3t^3\int{d^3\mathbf{r_0}d^3\mathbf{r}'\left[G_0^{\bot}(\mathbf{r}_0)\mathbf{P}(\mathbf{r}_0)+G_0^{\parallel}(\mathbf{r}_0)\mathbf{Q}(\mathbf{r}_0)\right]}\times \left[G_0^{\bot}(\mathbf{r}')\mathbf{P}(\mathbf{r}')+G_0^{\parallel}(\mathbf{r}')\mathbf{Q}(\mathbf{r}')\right]\\
&\times\exp{\left[i\mathbf{p}\left(\mathbf{r}_0+\mathbf{r}'\right)\right]}\left\{h_2(\mathbf{r}_0)h_2(\mathbf{r}')+h_2(\mathbf{r}_0+\mathbf{r}')\left[1+h_2(\mathbf{r}')\right]\left[1+h_2(\mathbf{r}_0)\right]\right\}\\
=&n_0^3t^3\left\{\int{d^3\mathbf{r_0}\left[G_0^{\bot}(\mathbf{r}_0)\mathbf{P}(\mathbf{r}_0)+G_0^{\parallel}(\mathbf{r}_0)\mathbf{Q}(\mathbf{r}_0)\right]h_2(\mathbf{r}_0)}\right\}^2\\
&+n_0^3t^3\int{d^3\mathbf{r_0}d^3\mathbf{r}'}h_2(\mathbf{r}_0+\mathbf{r}')\left[G_0^{\bot}(\mathbf{r}_0)\mathbf{P}(\mathbf{r}_0)+G_0^{\parallel}(\mathbf{r}_0)\mathbf{Q}(\mathbf{r}_0)\right]\left[1+h_2(\mathbf{r}_0)\right]\\
&\times\left[G_0^{\bot}(\mathbf{r}')\mathbf{P}(\mathbf{r}')+G_0^{\parallel}(\mathbf{r}')\mathbf{Q}(\mathbf{r}')\right]\exp{\left[i\mathbf{p}\left(\mathbf{r}_0+\mathbf{r}'\right)\right]}\left[1+h_2(\mathbf{r}')\right]
\end{split}
\end{equation}
To solve above equation, we resort to the Fourier transform of pair correlation function as \cite{Torquato1986PRB,tsang2004scattering2}
\begin{equation}
H_2(\mathbf{k})=\frac{1}{(2\pi)^3}\int d\mathbf{r}h_2(\mathbf{r})\exp{(-i\mathbf{k}\cdot\mathbf{r})}
\end{equation}
\begin{equation}
h_2(\mathbf{r})=\int d\mathbf{r}H_2(\mathbf{k})\exp{(i\mathbf{k}\cdot\mathbf{r})}
\end{equation}
Knowing that $Y_l^{m}(\theta,\phi)=\left(\frac{2l+1}{4\pi}\frac{(l-m)!}{(l+m)!}\right)^{1/2}P_l^m(\cos{\theta})\exp{(im\phi)}$ and $P_l^{-m}(\cos{\theta})=(-1)^m\frac{(l-m)!}{(l+m)!}P_l^m(\cos{\theta})$, we have
\begin{equation}
Y_l^{-m}(\theta,\phi)=\left(\frac{2l+1}{4\pi}\frac{(l+m)!}{(l-m)!}\right)^{1/2}P_l^{-m}(\cos{\theta})\exp{(-im\phi)}=(-1)^mY_l^{m*}(\theta,\phi)
\end{equation}
Using the property of spherical harmonics, we have
\begin{equation}
Y_l^m(0,\phi)=\sqrt{\frac{2l+1}{4\pi}}\delta_{m0}
\end{equation}
Using the well-known plane wave expansion \cite{abramowitz1964handbook}
\begin{equation}
\begin{split}
\exp{(i\mathbf{k}\cdot(\mathbf{r}_0+\mathbf{r}'))}=&\exp{(i\mathbf{k}\cdot\mathbf{r}_0)}\exp{(i\mathbf{k}\cdot\mathbf{r}')}\\
=&\left[4\pi\sum_{l,m}^{}i^lj_l(kr_0)Y_l^{m*}(\theta_0,\phi_0)Y_l^m(\theta_k,\phi_k)\right]\times\left[4\pi\sum_{l',m'}^{}i^lj_{l'}(kr')Y_{l'}^{m'}(\theta',\phi')Y_{l'}^{m'*}(\theta_k,\phi_k)\right]
\end{split}
\end{equation}
Thus we expand pair correlation function in its Fourier components as 
\begin{equation}
\begin{split}
h_2(\mathbf{r}_1+\mathbf{r}')=&\int{ d\mathbf{k}H_2(\mathbf{k})}\left[4\pi\sum_{l,m}^{}i^lj_l(kr_0)Y_l^{m*}(\theta_0,\phi_0)Y_l^m(\theta_k,\phi_k)\right]\\
&\times\left[4\pi\sum_{l,m}^{}i^lj_{l'}(kr')Y_{l'}^{m'}(\theta',\phi')Y_{l'}^{m'*}(\theta_k,\phi_k)\right]\\
=&(4\pi)^2\sum_{l,m}\sum_{l',m'}i^{l+l'}Y_l^{m*}(\theta_0,\phi_0)Y_{l'}^{m'}(\theta',\phi')\\
&\times\int k^2dkH_2(k)j_l(kr_1)j_{l'}(kr')\int d(\cos{\theta_k})d\phi_kY_l^m(\theta_k,\phi_k)Y_{l'}^{m'*}(\theta_k,\phi_k)\\
=&(4\pi)^2\sum_{l,m}\sum_{l',m'}i^{l+l'}Y_l^{m*}(\theta_0,\phi_0)Y_{l'}^{m'}(\theta',\phi')\int k^2dkH_2(k)j_l(kr_0)j_{l'}(kr')\delta_{ll'}\delta_{mm'}\\
=&(4\pi)^2\sum_{l,m}i^{2l}Y_l^{m*}(\theta_0,\phi_0)Y_{l}^{m}(\theta',\phi')\int k^2dkH_2(k)j_l(kr_0)j_{l'}(kr')
\end{split}
\end{equation}

In above equation, the normalization property of spherical harmonics is used. Thus the second term in $\mathbf{\Sigma}^{(3,PS)}$ is expressed as
\begin{equation}
\begin{split}
\mathbf{\Sigma}^{(3,PS,2)}(\omega,\mathbf{p})=&n_0^3t^3(4\pi)^2\sum_{l=1}^{N}\sum_{m=-l}^{m=l}(-1)^l\int k^2dkH_2(k)\int d^3\mathbf{r}_0d^3\mathbf{r}'j_l(kr_0)Y_l^{m*}(\theta_0,\phi_0)Y_l^m(\theta',\phi')\\
&\times\left\{\left[G_0^{\bot}(r_0)\mathbf{P}(\theta_0,\phi_0)+G_0^{\parallel}(r_0)\mathbf{Q}(\theta_0,\phi_0)\right]\left[1+h_2(r_0)\right]\exp{(ipr_0\cos{\theta_0})}\right\}\\
&\times\left\{\left[G_0^{\bot}(r')\mathbf{P}(\theta',\phi')+G_0^{\parallel}(r')\mathbf{Q}(\theta',\phi')\right]\left[1+h_2(r')\right]\exp{(ipr'\cos{\theta'})}\right\}
\end{split}
\end{equation}
Thus we have split the coupled integration into two integrals for $\mathbf{r}_0$ and $\mathbf{r'}$ which can be evaluated individually.
For convenience, let $ Y_l^{m}(\theta,\phi)=\left(\frac{2l+1}{4\pi}\frac{(l-m)!}{(l+m)!}\right)^{1/2}P_l^{m}(\cos{\theta})\exp{(im\phi)}=\gamma_l^mP_l^m(\cos{\theta})\exp{(im\phi)}$.
Moreover, it is noticed the integral involving the azimuth coordinate $\phi$ are zeros except for $m=-2,-1,0,1,2$, enabling us to integrate over $\phi$ first. After tedious but trivial algebra, we obtain
\begin{equation}
\begin{split}
\mathbf{\Sigma}^{(3,PS,2)}(\omega,\mathbf{p})=&n_0^3t^3(4\pi)^2\int k^2dkH_2(k)\sum_{l=0}^{N}(-1)^l\pi^2
\left(\begin{matrix}
&(2K_1+K_2)^2 &  & \\
& &(2K_1+K_2)^2 &  \\
&  & &(2K_1+2K_3)^2
\end{matrix}\right)\\
&+\sum_{l=1}^{N}(-1)^l\pi^2\left(
\begin{matrix}
&2K_4^2 &  & \\
& &2K_4^2 &  \\
&  & &4K_4^2
\end{matrix}\right)
+\sum_{l=2}^{N}(-1)^l\pi^2\left(
\begin{matrix}
&K_5^2 &  & \\
& &K_5^2 &  \\
&  & &0
\end{matrix}\right)
\end{split}
\end{equation}
where
\begin{equation}\notag
K_1=\int_{0}^{\infty} r^2dr G_0^{\bot}(r)j_l(kr)\left[1+h_2(r)\right]\gamma_l^0\int_{-1}^{1}d\mu P_l^0(\mu)\exp{(ipr\mu)}
\end{equation}
\begin{equation}\notag
K_2=\int_{0}^{\infty} r^2dr\left(G_0^{\parallel}(r)-G_0^{\bot}(r)\right)j_l(kr)\left[1+h_2(r)\right]\gamma_l^0\int_{-1}^{1}P_l^0(\mu)\exp{(ipr\mu)}(1-\mu^2)
\end{equation}
\begin{equation}\notag
K_3=\int_{0}^{\infty} r^2dr\left(G_0^{\parallel}(r)-G_0^{\bot}(r)\right)j_l(kr)\left[1+h_2(r)\right]\gamma_l^0\int_{-1}^{1}P_l^0(\mu)\exp{(ipr\mu)}\mu^2
\end{equation}
\begin{equation}\notag
K_4=\int_{0}^{\infty} r^2dr\left(G_0^{\parallel}(r)-G_0^{\bot}(r)\right)j_l(kr)\left[1+h_2(r)\right]\gamma_l^1\int_{-1}^{1}P_l^1(\mu)\exp{(ipr\mu)}\mu\sqrt{1-\mu^2}
\end{equation}
\begin{equation}\notag
K_5=\int_{0}^{\infty} r^2dr\left(G_0^{\parallel}(r)-G_0^{\bot}(r)\right)j_l(kr)\left[1+h_2(r)\right]\gamma_l^2\int_{-1}^{1}P_l^2(\mu)\exp{(ipr\mu)}(1-\mu^2)
\end{equation}
The high-order terms of $l$ in the infinite series involving integration over $\mu=\cos{\theta}$ is calculated through the recursive relation of associated Legendre functions $P_l^m(\mu)$ after the first several terms at $l=1,2$ in the series are calculated \cite{abramowitz1964handbook}. Afterwards, the integration over $r$ can be evaluated through numerical integrating methods easily without much computation load. We have checked $N=3$ is adequate to achieve convergence for above infinite sums.

Other terms in $\mathbf{\Sigma}^{(3)}(\omega,\mathbf{p})$ can be easily evaluated as:
\begin{equation}
\begin{split}
\mathbf{\Sigma}^{(3,PS,1)}(\omega,\mathbf{p})=&n_0^3t^3\left\{\int{d^3\mathbf{r}\left[G_0^{\bot}(\mathbf{r})\mathbf{P}(\mathbf{r})+G_0^{\parallel}(\mathbf{r})\mathbf{Q}(\mathbf{r})\right]h_2(\mathbf{r})}\exp{(i\mathbf{p}\cdot\mathbf{r})}\right\}^2\\
=&n_0^3t^3\left\{\int{r^2drh_2(r)\left[G_0^{\bot}(r)j_0(pr)\mathbf{I}+\left(G_0^{\parallel}(r)-G_0^{\bot}(r)\right)
\left(\begin{matrix}
&\frac{j_1(pr)}{pr} &  & \\
& &\frac{j_1(pr)}{pr} &  \\
&  & &\frac{j_0(pr)pr-j_1(pr)}{pr}
\end{matrix}\right)
\right]}\right\}^2
\end{split}
\end{equation}
The last part obtained from the integration of singular part of Green{'}s function is
\begin{equation}
\begin{split}
\mathbf{\Sigma}^{(3,sing)}(\omega,\mathbf{p})=-\frac{2 n_0^3t^3}{3k^2}\int d^3\mathbf{r}\mathbf{G}_0(\mathbf{r})h_2(\mathbf{r})\exp{(i\mathbf{p}\cdot\mathbf{r})}=-\frac{2 n_0t}{3k^2}\mathbf{\Sigma}^{(2,a)} (\omega,\mathbf{p})
\end{split}
\end{equation}
Finally, we have obtained 
\begin{equation}
\mathbf{\Sigma}^{(3)}(\omega,\mathbf{p})=\mathbf{\Sigma}^{(3,a)}(\omega,\mathbf{p})+\mathbf{\Sigma}^{(3,b)}(\omega,\mathbf{p})=\mathbf{\Sigma}^{(3,PS,1)}(\omega,\mathbf{p})+\mathbf{\Sigma}^{(3,PS,2)}(\omega,\mathbf{p})+\mathbf{\Sigma}^{(3,sing)}(\omega,\mathbf{p})
\end{equation}
\section{Derivation of pair correlation function $h_2(r)$ for hard-sphere and sticky-sphere systems}
We first calculate the pair correlation function in reciprocal space. For hard-sphere system, under the Percus-Yevick approximation, it is calculated simply as \cite{Baxter1968} 
\begin{equation}
H(\mathbf{p})=\frac{C(\mathbf{p})}{1-n_0(2\pi)^3C(\mathbf{p})}
\end{equation}
where
\begin{equation}
\begin{split}
C(\mathbf{p})=C(p)=&24f_v[\frac{\alpha+\beta+\delta}{u^2}\cos u-\frac{\alpha+2\beta+4\delta}{u^3}\sin u\\&-2\frac{\beta+6\delta}{u^4}\cos u+\frac{2\beta}{u^4}+\frac{24\delta}{u^6}(\cos u-1)]
\end{split}
\end{equation}
in which $u=4pa$, $\alpha=(1+2f_v)^2/(1-f_v)^4$, $\beta=-6f_v(1+f_v/2)^2/(1-f_v)^4$, $\delta=f_v(1+2f_v)^2/[2(1-f_v)^2]$ and $f_v$ is the volume fraction of the identical spherical particles.
Therefore, pair correlation function is obtained through Fourier transform into real space
\begin{equation}\label{ftransform}
h_2(\mathbf{r})=h_2(r)=\int_{-\infty}^{\infty}H(\mathbf{p})\exp{(i\mathbf{p}\cdot\mathbf{r})}d\mathbf{r}
\end{equation}

The sticky-sphere system is described by the following attractive potential as \cite{tsang2004scattering,Frenkel2002}, 
\begin{equation}
u(\mathbf{r})=\begin{cases}
\infty &{0<r<s}\\
\ln{[\frac{12\tau (b-s)}{b}]} & {s<r<b}\\
0 &{r>b}
\end{cases}
\end{equation}
where $\tau$ is the stickiness whose inverse quantifies the strength of inter-particle adhesion, $b=2a$ is particle diameter and $(b-s)$ stands for the range of potential, which is assumed to be infinitesimal because this potential is confined on particle surface. The Percus-Yevick approximation of the pair distribution function for the sticky hard spherical particles can be solved analytically using the factorization method of Baxter \cite{Baxter1968,tsang2004scattering2}.  Similarly, the pair correlation function in reciprocal space as
\begin{equation}
\begin{split}
[1+n_0(2\pi)^3H(\mathbf{p})]^{-1}=&\left\{\frac{f_v}{1-f_v}\left[\left(1-tf_v+\frac{3f_v}{1-f_v}\right)\Phi(x)+[3-t(1-f_v)]\Psi(x)\right]+\cos x\right\}^2\\
&+\left\{\frac{f_v}{1-f_v}[x\Phi(x)]+\sin x\right\}^2
\end{split}
\end{equation}
where $x=pa$, $\Psi(x)=3(\sin x/x^3-\cos x/x^2)$ and $\Phi(x)=\sin x/x$. And the parameter $t$ satisfies the following equation for a given $f_v$ and $\tau$ \cite{tsang2004scattering2}
\begin{equation}
\frac{f_v}{12}t^2-(\tau+\frac{f_v}{1-f_v})t+\frac{1+f_v/2}{(1-f_v)^2}=0
\end{equation}
Therefore, by taking inverse Fourier transform same as Eq. (\ref{ftransform}), the pair correlation function for sticky-sphere system is also obtained.
\section{Theoretical derivation of LDOS $\rho(\omega)$}
Now we give a brief derivation on LDOS, which can be calculated from averaged Green'{s} tensor as \cite{lagendijk1996resonant}
\begin{equation}
\begin{split}
\rho(\omega,\mathbf{R}=0)=&-\frac{2\omega}{\pi c^2}\mathrm{Im}[\mathrm{Tr}\mathbf{G}_c(\omega,\mathbf{R}=\mathbf{r}-\mathbf{r})]\\
=&-\frac{2\omega}{\pi c^2}\frac{1}{(2\pi)^3}\mathrm{Im}\left[\mathrm{Tr}\int{\mathbf{G}_{c}(\omega,\mathbf{p})d\mathbf{p}}\right]\\
\end{split}
\end{equation}

Since
\begin{equation}
\begin{split}
\int{\mathbf{G}_{c}(\omega,\mathbf{p})d\mathbf{p}}=&\int_0^{\infty}\left[\frac{\mathbf{I}-\mathbf{\hat{p}}\mathbf{\hat{p}}}{\epsilon^\bot(\omega,p)(\omega/c_0)^2-p^2}+\frac{\mathbf{\hat{p}}\mathbf{\hat{p}}}{\epsilon^\parallel(\omega,p)(\omega/c_0)^2}\right]d\mathbf{p}\\
=&2\pi \mathbf{I}\int_{-\infty}^{\infty}{\left[ \frac{2}{3}\frac{1}{\epsilon ^{\bot}\left( \omega ,p \right) \left( \omega /c \right) ^2-p^2}+\frac{1}{3}\frac{1}{\epsilon ^{\parallel}\left( \omega ,p \right) \left( \omega /c \right) ^2} \right]}p^2dp
\\	
\end{split}
\end{equation}
Using contour integration technique, above equation can be evaluated through residual theorem as
\begin{equation}
\begin{split}
\int{\mathbf{G}_{c}(\omega,\mathbf{p})d\mathbf{p}}=&
2\pi \mathbf{I}\left[ -\frac{2\pi iK^{\bot}}{3}+\int_{-\infty}^{\infty}{\frac{1}{3}\frac{p^2dp}{\epsilon ^{\parallel}\left( \omega ,p \right) \left( \omega /c \right) ^2}} \right]\\	
\end{split}
\end{equation}
Finally, we obtain
\begin{equation}
\begin{split}
\rho(\omega,\mathbf{R}=0)=&\frac{n^{\bot}\omega^2}{\pi^2 c^3}-\frac{1}{2\pi^3\omega}\mathrm{Im}\int{\frac{p^2dp}{\epsilon^{\parallel}(\omega,p)}}
\end{split}
\end{equation}
where $K^\bot=n^\bot\omega/c$ is the effective propagation constant of the transverse mode.
\section{Coupled-dipole approach to numerically calculate LDOS}
To implement validation for the previous theoretical formulations, here we present coupled-dipole approach (CDA) as an exact solution for dipole scatterers distributed in a spherical volume to calculate the exact LDOS in classical electrodynamics. The coupled-dipole equations are \cite{Pierrat2010,Caze2013}
\begin{equation}\label{coupled-dipole}
\mathbf{d}_j(\omega)=\frac{\omega^2\alpha(\omega)}{c^2}\left[\mathbf{G}_0(\omega,\mathbf{r}_j,\mathbf{r}_s)\mathbf{d}_s+\sum_{i=1,i\neq j}^{N}\mathbf{G}_0(\omega,\mathbf{r}_j,\mathbf{r}_i)\mathbf{d}_i(\omega)\right]
\end{equation}
where $\mathbf{d}_s$  is the dipole moment of the emitting source and $\mathbf{d}_j(\omega)$ is the excited dipole moment of the $i$-th particle. $\mathbf{G}_{0}(\omega,\mathbf{r}_j,\mathbf{r}_s)$ is the free-space dyadic Green's function describing emitting field from the source $\mathbf{r}_s$ to the $i$-th dipole and   $\mathbf{G}_{0}(\omega,\mathbf{r}_j,\mathbf{r}_i)$  describes the propagation of scattering field of $j$-th dipole to $i$-th dipole. 
%For the convenience of calculation, here we adopt a slightly different form of free-space dyadic Green’s function as
%\begin{equation}
%\begin{split}
%\tilde{\mathbf{G}}_{0}(\omega,\mathbf{r}_j,\mathbf{r}_i)=&\frac{\exp{(ikr)}}{4\pi r}\left(\frac{i}{kr}-\frac{1}{k^2r^2}+1\right)\mathbf{I}\\
%&+\frac{\exp{(ikr)}}{4\pi r}\left(-\frac{3i}{kr}+\frac{3}{k^2r^2}-1\right)\mathbf{\hat{r}}\mathbf{\hat{r}}
%\end{split}
%\end{equation}
where the singular part is eliminated due to the fact that all dipoles will not overlap at all. After calculating the EM responses of all scatterers based on above multiple scattering equations, we find the total scattering field of the random cluster of particles at an arbitrary position $\mathbf{r}\neq\mathbf{r}_j$ where $\mathbf{r}_j$ denotes the positions of scatterers, is computed as
\begin{equation}
\mathbf{E}_s(\mathbf{r})=\frac{\omega^2}{c^2}\sum_{i=1}^{N}\mathbf{G}_0(\omega,\mathbf{r},\mathbf{r}_i)\mathbf{d}_i(\omega)
\end{equation}
From the scattering fields, it is straightforward to obtain the full Green's function of the disordered medium as $\mathbf{G}_c(\mathbf{r},\mathbf{r}_s)=\mathbf{G}_0(\mathbf{r},\mathbf{r}_s)+\mathbf{S}(\mathbf{r},\mathbf{r}_s)$ for a point emitter (source) placed at $\mathbf{r}_s$  after performing ensemble average over a large amount of configurations. The subscript $c$ actually denotes a configurational average procedure.  Here the elements in scattering field tensor $\mathbf{S}(\mathbf{r},\mathbf{r}_s)$ can be calculated through the relation $\mathbf{E}_s(\mathbf{r})=\mathbf{S}(\mathbf{r},\mathbf{r}_s)\mathbf{d}_s$, by aligning the dipole moment with different axes. Afterwards LDOS is obtained from the ensemble averaged Green’s function as
\begin{equation}
\rho(\mathbf{r},\omega)=\frac{2\omega}{\pi c^2}\mathrm{Im}\left[\mathrm{Tr}\mathbf{G}_c(\mathbf{r}_s,\mathbf{r}_s)\right]
\end{equation}
\bibliography{nfi2018}